\documentclass[bibyear]{aa}  

\usepackage{graphicx}

\usepackage{txfonts}
\usepackage{booktabs}
\usepackage{siunitx}
\usepackage{comment}
\usepackage{caption} 
\usepackage{natbib}
\bibpunct{(}{)}{;}{a}{}{,} 

\usepackage[colorlinks]{hyperref}
\usepackage{threeparttable}

\hypersetup{
    colorlinks = True,
    citecolor = blue
}

\newcommand{\orcit}[1]{\protect\href{https://orcid.org/#1}{\protect\includegraphics[width=8pt]{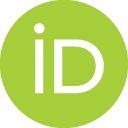}}}

\newcommand{\msun}{\ensuremath{\mathrm{M}_\odot}}

\usepackage{xspace}
\newcommand{\ruwe}{\ifmmode\mathtt{RUWE}\else\texttt{RUWE}\fi\xspace}

\newcommand{\Eone}{\href{https://simbad.cds.unistra.fr/simbad/sim-id?Ident=Gaia+DR3+3669003373413914624}{E1}\xspace}
\newcommand{\Etwo}{\href{https://simbad.cds.unistra.fr/simbad/sim-id?Ident=Gaia+DR3+4073896567545411328}{E2}\xspace}
\newcommand{\Ethree}{\href{https://simbad.cds.unistra.fr/simbad/sim-id?Ident=Gaia+DR3+4285576371518811776}{E3}\xspace}
\newcommand{\Efour}{\href{https://simbad.cds.unistra.fr/simbad/sim-id?Ident=Gaia+DR3+4207618935502749184}{E4}\xspace}

\newcommand{\Sone}{\href{https://simbad.cds.unistra.fr/simbad/sim-id?Ident=Gaia+DR3+1193137104465445888}{S1}\xspace}
\newcommand{\Stwo}{\href{https://simbad.cds.unistra.fr/simbad/sim-id?Ident=Gaia+DR3+6771307454464848768}{S2}\xspace}
\newcommand{\Sthree}{\href{https://simbad.cds.unistra.fr/simbad/sim-id?Ident=Gaia+DR3+6553439603373054720}{S3}\xspace}
\newcommand{\Sfour}{\href{https://simbad.cds.unistra.fr/simbad/sim-id?Ident=Gaia+DR3+6032843799994897408}{S4}\xspace}
\newcommand{\Sfive}{\href{https://simbad.cds.unistra.fr/simbad/sim-id?Ident=Gaia+DR3+5764934181067920256}{S5}\xspace}

\newcommand{\Oone}{\href{https://simbad.cds.unistra.fr/simbad/sim-id?Ident=Gaia+DR3+5239771349915805952}{O1}\xspace}

\newcommand{\Aone}{\href{https://simbad.cds.unistra.fr/simbad/sim-id?Ident=Gaia+DR3+491104466350124544}{A1}\xspace}

\newcommand{\rev}[1]{#1}

\begin{document}

   \title{Hide and Seek with Gaia: detectability of Predicted Thin-Disc Metal-Rich RR Lyrae  Binaries in Gaia DR3 and DR4.}

   \subtitle{}

   \author{Giuliano Iorio
           \orcit{0000-0003-0293-503X}\inst{1}\thanks{\href{mailto:giuliano.iorio.astro@gmail.com}{giuliano.iorio.astro@gmail.com}}
           \and
           Pranav Nagarajan\orcit{0000-0002-1386-0603}\inst{2}
           \and
           Alexey Bobrick\orcit{0000-0002-4674-0704}\inst{3,4}
           \and
           Kareem El-Badry\orcit{0000-0002-6871-1752}\inst{2}
           \and
           Elena Pancino\orcit{0000-0003-0788-5879}\inst{5}
           \and
           Vasily Belokurov\inst{6}
           \and 
           HanYuan Zhāng\orcit{0009-0005-6898-0927}\inst{6}
           \and
           Valentina D'Orazi\orcit{0000-0002-2662-3762}\inst{7,8}
           \and
           Cecilia Mateu\orcit{0000-0002-6330-2394}\inst{9}
           \and 
           Sara Rastello\orcit{0000-0002-5699-5516}\inst{1,10}
           \and
           Mark Gieles\orcit{0000-0002-9716-1868}$^{1,11,12}$
          }

\institute{
Institut de Ciències del Cosmos (ICCUB), Universitat de Barcelona (UB), c. Martí i Franquès 1, 08028 Barcelona, Spain
\and
Department of Astronomy, California Institute of Technology, 1200 E. California Blvd., Pasadena, CA 91125, USA
\and
School of Physics and Astronomy, Monash University, Clayton, Victoria 3800, Australia
\and
ARC Centre of Excellence for Gravitational Wave Discovery -- OzGrav, Australia
\and
INAF -- Osservatorio Astrofisico di Arcetri, Largo Fermi 5, 50125 Firenze, Italy
\and
Institute of Astronomy, University of Cambridge, Madingley Road, Cambridge CB3 0HA, UK
\and
Department of Physics, University of Rome Tor Vergata, via della Ricerca Scientifica 1, 00133, Rome, Italy
\and
INAF Osservatorio Astronomico di Roma, via Frascati 33, Monte Porzio Catone, Italy
\and 
Departamento de Astronom\'{i}a, Instituto de F\'{i}sica, Universidad de la Rep\'{u}blica, Igu\'a 4225, CP 11400 Montevideo, Uruguay
\and
Departament de Física Quàntica i Astrofísica (FQA), Universitat de Barcelona (UB), c. Martí i Franquès 1, 08028 Barcelona, Spain
\and
   ICREA, Pg. Llu\'is Companys 23, 08010 Barcelona, Spain
\and 
Institut d’Estudis Espacials de Catalunya (IEEC), Edifici RDIT, Campus UPC, 08860 Castelldefels (Barcelona), Spain
}

   \date{}

\abstract{
RR Lyrae stars (RRLs) are classical tracers of old stellar populations, yet growing evidence suggests the presence of a metal-rich ($\mathrm{[Fe/H]}\gtrsim-0.5$), intermediate-age (2--7 Gyrs) sub-population in the Milky Way disc. Binary evolution, particularly stable mass transfer, has been proposed as a viable formation channel, predicting that most metal-rich ($\mathrm{[Fe/H]}\gtrsim-0.5$), intermediate-age ($t_{\mathrm{age}}\lesssim9$ Gyr) RRLs should reside in binaries with orbital periods of $\sim$900–2000 days. 
However, no genuine RRL binaries have been robustly identified, including in the Gaia DR3 astrometric binary catalogues, despite Gaia being sensitive to the predicted orbital-period range.
We investigate whether the lack of detections in Gaia DR3 reflects an intrinsically low binary fraction or instead arises from observational biases. We analyse a carefully selected sample of 100 Gaia DR3 RRLs designed to trace the metal-rich population with thin-disc kinematics and compare them with predictions from binary evolution models. We generate realistic Gaia observation mocks, including variability-induced astrometric biases, and assess the detectability of binaries and the posterior constraints on the hidden binary fraction using astrometric quality indicators, such as \ruwe, and a robust Bayesian inference.
While current uncertainties prevent a definitive rejection of a high fraction of hidden binaries, our results reveal tensions between existing binary evolution predictions and the Gaia DR3 non-detections. This suggests either the presence of unaccounted systematics in the modelling of Gaia observations or the need to revise assumptions in binary evolution models.  We predict that Gaia DR4 will significantly improve the binary detectability and provide powerful new constraints on the post-interaction binary populations.
}

   \keywords{}

   \titlerunning{Detectability of predicted RR Lyrae binaries in Gaia}
   \authorrunning{G. Iorio et al.}

   \maketitle

\section{Introduction}

RR Lyrae stars  (RRLs) are short-period pulsating variables ($0.2 \lesssim P \lesssim 1$~days) exhibiting radial oscillations in the fundamental (RRab), first-overtone (RRc), or both modes simultaneously (RRd). 
From an evolutionary standpoint, they are low-mass stars that have undergone the helium flash and are currently in the core-helium-burning phase crossing the instability strip in the Hertzsprung--Russell diagram with effective temperature in the range $T_{\rm eff}\approx 5500-8000$~K, and luminosity in the range $L\approx 30-90 \ \mathrm{L}_\odot$ \citep{Smolec08,Catelan15,DeSomma24}.
Stellar evolution models predict that the formation of RRLs is most efficient for metal-poor ([Fe/H]$\sim -1.5$) and low-mass ($\approx 0.8$--$1.0\,\msun$) progenitors, characteristic of Population~II stars with ages $t_\mathrm{age} \gtrsim 10$~Gyr. Accordingly, RRLs are abundant in old and metal-poor stellar populations (e.g. the Galactic halo and globular clusters, see e.g. \citealt{Reyes24,Alonso25}).

The astrophysical relevance of RRLs stems from the well-established theoretical and empirical correlations between their light-curve properties and intrinsic stellar parameters, such as luminosity, pulsation period,  and metallicity \citep[see e.g.][]{Dekany22,Mullen23,Narloch24,Prudil24,Lengen25,Monti25,Muraveva25}. Their relatively high luminosity (typical absolute visual magnitude $M_V \simeq 0.5$), narrow range of colours, and distinctive light-curve morphology ensure a low level of contamination \citep[see e.g.][]{Iorio18}, making RRLs reliable standard candles and powerful tracers of old and metal-poor stellar populations and Galactic structure in the Milky Way and its nearby satellites \citep[see e.g.][]{Belokurov18,IB19,Price19,Nagarajan22,Savino22,Ablimit22,Kunder22,Garofalo24,Han25,Bono26}.

Despite the robust understanding of the evolutionary pathways of their progenitors and of the pulsation mechanism, a few key questions remain still open and debated. Among these are the origin of metal-rich  RRLs ([Fe/H]$\gtrsim -1.0$), the existence of young RRLs ($t_\mathrm{age} < 10~\mathrm{Gyr}$), as well as the apparent scarcity of RRLs in binary systems. 
Investigating the formation channels of metal-rich RRLs and their connection to the apparent scarcity of RRL binaries is central to our understanding of RRL formation and population properties. Moreover, it has important implications for stellar evolution and binary interaction processes, such as mass transfer and mergers, as well as for the interpretation of data from current and forthcoming large-scale surveys (see, e.g., \citealt{Ibata23} and the discussion in \citealt{Zhang25}). 

\section{Metal-rich and/or young RR Lyrae stars }

\begin{figure}
    \centering
    \includegraphics[width=\linewidth]{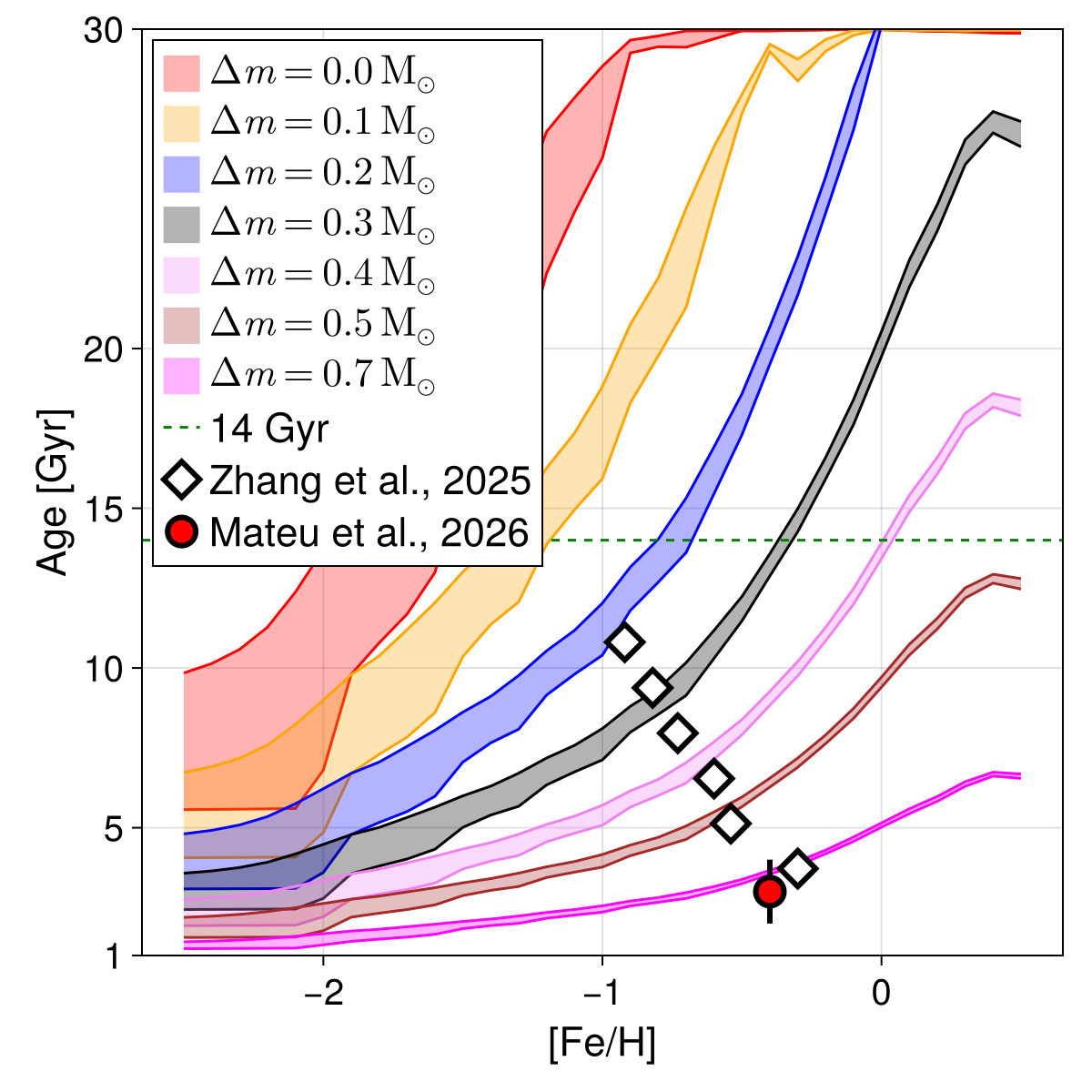}
    \caption{Expected age range of RRLs based on single star models (shaded areas as a function of metallicity and progenitor mass loss, $\Delta m$). The models are based on the solar-scaled PARSEC v1.2S stellar tracks \citep{Chen15} and are computed following the method described in Appendix~C of \cite{Zhang25}, from which we also take the age estimates of RRL populations inferred via a kinematic comparison with O-rich Mira variable stars (white diamonds). The red dots show the metallicity and age range (2--4 Gyr) for the RRL associated with the Trumpler 5 open cluster \citep{Mateu25}. This figure shows that  high mass loss  is required to explain RRLs for [Fe/H]$\gtrsim-1$ and  age $< 5$ Gyr.}
    \label{fig:rrlage}
\end{figure}

The existence of a sub-population of metal-rich RRLs ($-1.0 \lesssim \mathrm{[Fe/H]} \lesssim 0$) in the Solar neighbourhood has been recognised since the pioneering works of \citet{Preston59} and \citet{Tam76}. Subsequent advances in astrometric and spectroscopic surveys have enabled a more detailed characterisation of this population \citep{Soszynski16,Crestani21,DOrazi24,Prudil25}, while the Gaia mission has provided the first homogeneous, all-sky view of metal-rich disc-like RRLs in the Milky Way \citep{Clementini23,IB21}.
Most Gaia RRLs have only photometric metallicities, estimated from correlations with their light-curve properties in the $G$-band (see e.g. \citealt{IB21,Dekany22,Muraveva25}). Although these estimates are significantly more uncertain than spectroscopic measurements and should be treated cautiously on a star-by-star basis, they capture population-level trends \citep{Clementini23}.
In this context, Gaia shows that the most metal-rich RRLs appear preferentially distributed in a thin-disc-like configuration and exhibit cold kinematics, consistent with those of intermediate-age stellar populations ($t_\mathrm{age} \approx 3$--$9$~Gyr; \citealt{Layden94,Prudil20,Zinn20,IB21,Cabrera24warp}). The population is alpha-poor, consistent with a disc population \citep[see e.g.][]{Marsakov18}, but shows peculiar under-abundances in some elements (yttrium, scandium, barium,  \citealt{Clementini95,Chadid17,Gozha21,Gozha24,DOrazi24}), qualitatively similar to the chemical patterns seen in some post-RGB objects 
(Molina et al. in prep., see also \citealt{Martin25,Mohorian25}).

\citet{Zhang25} compared the kinematics of RRLs with those of O-rich Mira variables, whose ages can be independently estimated from their pulsation periods, and identified a metallicity-dependent gradient in the kinematically inferred ages of RRLs, with younger objects being systematically more metal-rich (see Figure \ref{fig:rrlage}). In particular, they found that RRLs with $\mathrm{[Fe/H]} \gtrsim -0.5$ are predominantly associated with intermediate-to-young disc populations, with characteristic ages $t_\mathrm{age} \lesssim 5$~Gyr.
Direct evidence for the presence of RRLs in Population~I environments was reported by \citet{Mateu25}, who identified an RRL associated with the metal-rich ([Fe/H]$\approx -0.4$), intermediate-age (2–4~Gyr) open cluster Trumpler~5 \citep[][see Figure \ref{fig:rrlage}]{Donati15,Magrini23}. Although Trumpler~5 is often classified as an old and metal-poor open cluster (relative to other open clusters), it is significantly more metal-rich and younger than the bulk of the halo RRL population. This detection represents the first direct evidence that RRLs can also form in Population~I environments. Further studies have reported evidence for a significant population of RRLs in the Magellanic Clouds whose inferred age distribution is consistent with values spanning $\sim 1$–$8$~Gyr \citep{Cuevas24}.

As already noted by \citet{Tam76}, the existence of metal-rich RRLs with intermediate or young ages is paradoxical, since their higher atmospheric opacity would instead favour residence in the red clump during the core-helium-burning phase. The formation of metal-rich and/or young RRLs therefore requires alternative explanations, such as a substantial enhancement of mass loss during the red giant branch phase ($\gtrsim 0.5\,\msun$; Figure \ref{fig:rrlage}, see also  \citealt{Bono97a}). In this scenario, only a small hydrogen-rich envelope ($\sim 0.03-0.07\,\msun$) remains at the onset of core-helium burning, exposing the hot core and allowing the star to enter the instability strip. However, the mass-loss rates required by this mechanism, at least if they are mediated through stellar winds, are significantly larger than those typically inferred for red giant branch winds in dwarf galaxies, globular clusters and field stars ($\lesssim 0.3\,\msun$; see e.g. \citealt{Savino19,Tailo20,Brogaard24}).
\rev{In particular, although studies of RGB winds in the metal-poor regime of globular clusters indicate increasing efficiency with metallicity (see e.g.\ \citealt{Howell25}), there is growing evidence that around [Fe/H]$\approx-1$ this trend reverses, making the wind mass-loss efficiency progressively smaller with increasing metallicity (see e.g.\ \citealt{Yaguang25,Howell26}). The production of metal-rich RRLs through standard wind mass loss thus appears unlikely.}

\subsection{RR Lyrae stars in binary systems}

An alternative formation channel for metal-rich and/or young RRLs has been proposed in the form of mass transfer from a red giant (the RRL progenitor overfilling its Roche-lobe) to a binary companion. 
Roche-lobe overflow is substantially more efficient than wind mass loss. It can strip sufficient envelope mass ($\gtrsim 0.4,\msun$, see Figure~\ref{fig:rrlage}) to produce core-helium-burning stars hot enough to enter the instability strip even at high metallicity \citep{Karczmarek17,BI24}.

In particular, \citet{BI24} combined detailed binary evolution simulations performed with MESA \citep{Vos20} with a realistic model of the Milky Way stellar populations \citep{Robin03}, and showed that this scenario predicts the existence of metal-rich and intermediate-age RRLs in the Milky Way disc with properties consistent with those of the observed population.
A robust prediction of these models is that most metal-rich RRLs, and essentially all young ones ($t_\mathrm{age}<9$ Gyr), reside today in low-mass binary systems with orbital periods of $\sim 900$--$2000$~days. Moreover, this study also highlighted that nearly one-half of classical metal-poor RRLs also robustly have a binary companion (that they did not interact with) in the range of $500$--$4\times 10^9$~days.

Despite these theoretical expectations, both classical and metal-rich RRLs are observed to exhibit an exceptionally low incidence of binarity. Only one confirmed case of a short-period ($\approx15$~days) binary system hosting an RRL-like pulsator is known. This object, however, is not a genuine RRL, but a so-called binary evolutionary pulsator (BEP): a low-mass ($\approx 0.26\,\msun$) stripped post-RGB star crossing the instability strip on its way toward the white-dwarf cooling sequence \citep{BEP,Smolec13}.

Aside from this case, the majority of proposed RRL binary candidates have been identified indirectly, primarily through the Light Travel Time Effect  (LTTE), which manifests as periodic modulations in the pulsation phase induced by orbital motion \citep[$\sim 200$ with periods $\sim 10^2-10^4$ days, e.g.][]{Liska16,Hajdu21,Li21,Poretti2025}. 
\rev{Additional candidates have been reported using speckle interferometry \citep[$10$, with separations $\gtrsim 20$ AU,][]{Salinas26}} and proper-motion anomalies \citep[$\sim 200$,][]{Kervella19a,Kervella19b}. A statistical comparison of the properties of these candidates and with those of binary-formed RRLs predicted by the models does not reveal tensions \citep[see the discussion in ][]{BI24}.
Additionally, \cite{Hajdu26} finds that $\sim$1\% of RRLs in the bulge field show mean-magnitude variations, which could be interpreted, among other possibilities, as indicative of circumbinary material.
However, dedicated LTTE studies for thin-disc RRLs have so far yielded no candidates with orbital periods in the range predicted by binary evolution models \citep{Abdollahi2025}.
As possible indirect evidence of binarity, three s-process–enhanced RRLs are known \citep{Preston06,Dorazi25}, whose chemical enrichment likely requires past mass transfer from an evolved AGB companion. However, none of these sources has been independently confirmed as a binary system.

The apparent scarcity of binary systems among RRLs, particularly for the metal-rich population in the Milky Way disc, is not only in tension with the predictions of binary evolution models, but is also at odds with the significantly higher binary fractions observed among their progenitors, namely solar-mass main-sequence stars and red giants ($\sim 20$–$50\%$; see \citealt{Moe19}). 
For example, a comparable fraction  of classic metal-poor RRLs should have a companion that did not interact with the RRL progenitor. 
This discrepancy could reflect intrinsic physical effects related to binary evolution and mass-loss processes, potentially requiring a revision of current models, or observational biases affecting binary detection, or a combination of both.

In this context, it is important to note that observational biases and uncertainties are expected to have a significant impact on the detections of RRLs in binaries.
For example, the interpretation of LTTE detections remains uncertain, as this technique requires long observational baselines and is affected by degeneracies \rev{with other sources of phase modulation, most notably the Blazhko effect \citep[see e.g.][]{Sylla2024}, and possible mean-magnitude variations due to circumstellar or circumbinary material \citep[see e.g.][]{Hajdu26}.}
Spectroscopic detection of RRL binaries is likewise challenging, owing to the large intrinsic radial-velocity variations induced by pulsations (up to about $80~\mathrm{km\,s^{-1}}$), which can obscure the comparatively small orbital signals expected for long-period, low-mass companions (typically $\lesssim 10~\mathrm{km\,s^{-1}}$; \citealt{BI24}). To date, the only dedicated spectroscopic campaign targeting RRL binarity has not reached conclusive results \citep{Barnes21}.

In light of these limitations, astrometry, especially with current and forthcoming Gaia data releases, represents the most promising method for detecting the population predicted by these models.

\begin{figure*}
    \centering
    \includegraphics[width=\linewidth]{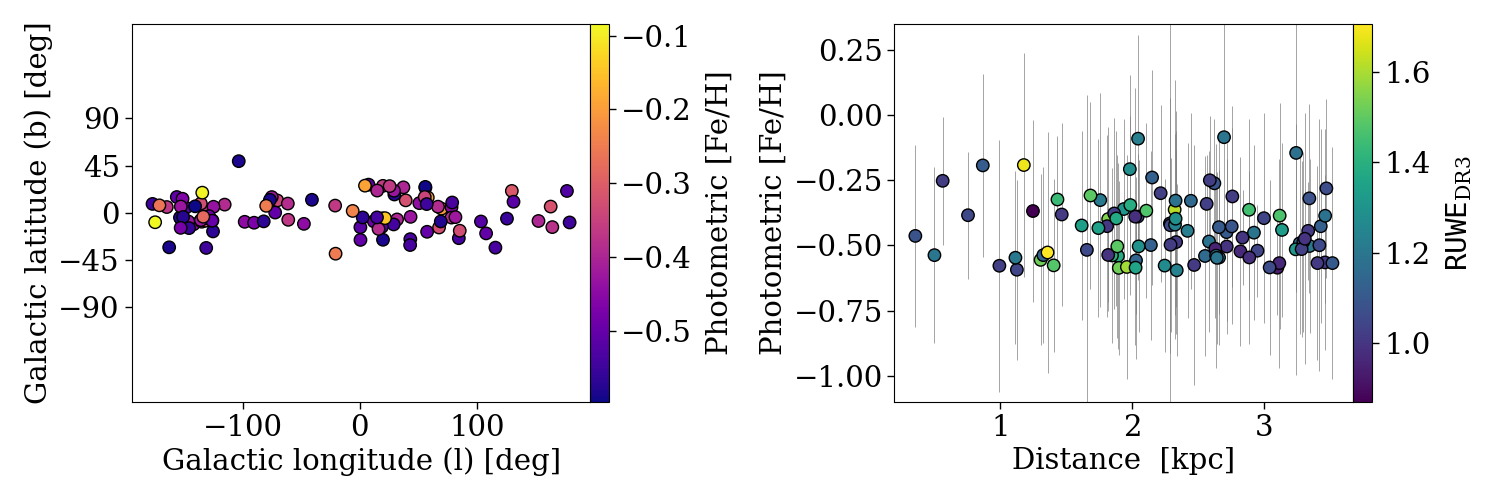}
    \caption{Properties of the Gaia DR3 RRL subsample used in our analysis.
Left: sky distribution in Galactic coordinates with colormap showing the photometric metallicity.
Right: photometric metallicity as a function of distance, with associated uncertainties shown as grey error bars and with colormap showing the Gaia DR3 \ruwe (see text). Distances are estimated as the inverse parallax after correcting for the global Gaia DR3 parallax zero-point offset.}
    \label{fig:rrlsample}
\end{figure*}

\subsection{This work: constraining the binary formation channel with Gaia DR3 RR Lyrae stars}

The Gaia mission has opened a new window for detecting binary systems through astrometric measurements \citep{Gaiamission,GaiaDR3}, either via excess residuals in single-star solutions \citep[e.g.][]{Belokurov20,Penoyre22} or, since Gaia DR3, through direct astrometric binary fits \citep{GaiaDR3binary}. Gaia DR3 is, in principle, particularly sensitive to the orbital-period range ($\sim$900–2000 days) predicted for RRLs formed through binary evolution \citep{Castro24,ElBadry24}, making it a powerful tool for testing such formation scenarios.

Among the $\sim2\times10^{5}$ RRLs identified in Gaia DR3 \citep{Clementini23}, only eleven appear in the astrometric binary catalogue, and none are genuine RRLs. These sources are instead misclassified as eclipsing binaries or other systems (Appendix \ref{app:gaiarrlbin}; see also \citealt{Pranav26}). While this might suggest a negligible binary fraction among RRLs, Gaia’s sensitivity is limited to specific regions of parameter space set by distance, orbital period, and astrometric signal-to-noise \citep[e.g.][]{GaiaDR3binary,ElBadry24mock}. In addition, intrinsic stellar variability may further bias astrometric binary detection \citep[e.g.][]{Belokurov20}.

In this work, we assess whether the absence of confirmed Gaia DR3 detections is statistically compatible with a high fraction of binaries with the properties predicted by the metal-rich RRL binary-formation model of \citet{BI24}. We combine a carefully selected sample of metal-rich Gaia DR3 RRLs that are kinematically consistent with intermediate-young thin-disc populations with forward modelling of the expected binary population. We then assess their detectability in Gaia DR3 and DR4 using an updated version of the \texttt{gaiamock} tool \citep{ElBadry24mock}, which explicitly accounts for variability-induced astrometric effects. The resulting mock observations are used to constrain the allowed binary fraction through both the predicted astrometric-noise properties and a Bayesian inference framework.

This study is complementary to the analysis presented in \cite{Pranav26}, which combines the same binary evolution simulations from \citet{BI24} with Galactic population models and \texttt{gaiamock} to predict the global population of stripped and partially stripped core-helium-burning stars, including but not limited to metal-rich RRLs, observable in Gaia DR3 and Gaia DR4. Here, we instead focus specifically on the metal-rich RRL population, adopting a more data-driven approach that accounts for variability-induced astrometric biases and directly constrains the posterior probability of a binary fraction that could remain hidden in the Gaia DR3 observations.

The paper is organised as follows. Section~\ref{sec:method} describes the methodology. The results are presented in Section~\ref{sec:results}. Section~\ref{sec:discussion} discusses the implications of our findings, including systematics, predictions for Gaia DR4, and a detailed comparison with the companion study by \cite{Pranav26}. Finally, Section~\ref{sec:summary} summarises our conclusions.

\section{Method} \label{sec:method}

We combine forward modelling with data-driven constraints, following a strategy similar to that of \cite{Culpan25}. We first construct a gold sample of bona fide, preferentially metal-rich, intermediate-age thin-disc RRLs from Gaia~DR3, adopting their observed positions, distances, and magnitudes as fixed inputs. Each RRL is then assumed to reside in a binary system with properties drawn from the formation models of \cite{BI24}, including orbital period, component masses, and $G$-band luminosity ratio. We subsequently simulate Gaia epoch astrometry and astrometric model fitting using the \texttt{gaiamock} tool \citep{ElBadry24mock}, updated to account explicitly for variability-induced astrometric effects \citep{Lindegren21,Lindegren22}. Model uncertainties and parameter degeneracies are explored via bootstrap resampling, generating 1000 binary realisations per source.

\subsection{The RR Lyrae sample} \label{sec:sample}

We select a subsample of bona fide RRLs from the Gaia DR3 catalogue \citep{Clementini23} designed 
to maximise the fraction of binary-formed RRLs by targeting metal-rich objects with thin-disc–like 
kinematics consistent with the intermediate-young disc population \citep{IB21,BI24,Zhang25}.
To this end, we use the same catalogue as in \citet{Zhang25}, consisting of a subsample of Gaia DR3 
RRLs derived from the clean RRL sample in \citet{IB21} and complemented with photometric 
metallicities following the methodology of \citet{IB21}. Photometric metallicities are derived from 
light-curve Fourier parameters calibrated on the \citet{Zinn84} scale \citep[see][]{IB21,Muraveva25}.

To produce the final sample, we apply the following additional selection criteria:
\begin{enumerate}
    \item parallax signal-to-noise ratio $(\varpi/\sigma_\varpi) > 2$;
    \item parallax $\varpi > 0.25$ mas;
    \item $\mathrm{[Fe/H]} > -0.6$.
\end{enumerate}

The parallax-based cuts select nearby sources ($\lesssim 4$ kpc; Section~\ref{sec:mock}), for which 
Gaia DR3 is sensitive to astrometric binarity \citep{GaiaDR3binary,ElBadry24mock}. The cut on 
parallax signal-to-noise ratio reduces the number of sources consistent with zero or negative 
parallaxes without imposing overly restrictive astrometric quality cuts. In our sample, 95\% of 
sources satisfy $(\varpi/\sigma_\varpi) > 3$, and 85\% satisfy $(\varpi/\sigma_\varpi) > 5$.
The metallicity cut favours the inclusion of intermediate-age thin-disc RRLs expected to 
preferentially (or exclusively) originate from binary evolutionary channels. In particular, most 
binary-formed RRLs in the model of \citet{BI24} have $\mathrm{[Fe/H]} > -0.5$ (see 
Appendix~\ref{app:DPGMM}), and \citet{Zhang25} showed that the population with 
$\mathrm{[Fe/H]} > -0.6$ captures the bulk of RRLs with kinematics consistent with intermediate-age 
disc populations ($\lesssim 7$--8 Gyr; see their Figure 7).

The resulting sample consists of 100 RRLs, with no spurious astrometric binaries 
(see Appendix~\ref{app:gaiarrlbin}). Their sky distribution, distances, photometric metallicities, 
and \ruwe\ (Renormalised Unit Weight Error) are shown in Figure~\ref{fig:rrlsample}. The \ruwe\ 
provides a diagnostic of the quality of the astrometric solution: values of $\ruwe \approx 1$ are 
expected for well-behaved solutions, whereas $\ruwe > 1.4$ typically indicates excess astrometric 
scatter, which may be associated with binarity, though not exclusively \citep[see 
e.g.][]{Lindegren18,Belokurov20,Castro24}.

Our sample is based on the clean RRL catalogue of \citet{IB21}, which applies quality cuts on 
extinction, colour excess, and DR2-based \ruwe, and removes Galactic substructures (e.g.\ stellar 
clusters and streams). While these cuts bias the sample toward well-behaved astrometric solutions, 
our goal is not to directly identify binaries, but to test whether a population of binaries predicted 
by the models could remain hidden in Gaia data. When considering the DR3-based \ruwe, the final 
sample nonetheless extends up to $\ruwe \approx 1.8$; tests using alternative selections are presented 
in Appendix~\ref{app:alternative} and discussed in Section~\ref{sec:disc:cuts}.

Figure~\ref{fig:rrlsample} shows a possible trend of \ruwe\ with distance, with the highest values 
more concentrated within 1--2.5 kpc. This high-\ruwe\ population is mostly composed of HASP 
(high-amplitude short-period) RRLs \citep[see e.g.][]{IB21}. However, given the strong spatial 
inhomogeneity introduced by the \citet{IB21} selection cuts (see their Figures 11 and 13 for the 
HASP population), this trend is more likely due to a combination of significant colour variation in 
HASP RRLs (see Section~\ref{sec:ruwe}) convolved with the Gaia scanning law and inhomogeneities 
in the selection function. We therefore refrain from drawing any conclusions about a distance trend.

Finally, we note that our sample may include some contamination from non-thin-disc populations. 
Given the large uncertainties in photometric metallicities, the statistical nature of the kinematic 
analysis in \citet{Zhang25}, and the possible kinematic overlap between the thin and thick discs 
\citep[see e.g.][]{Schonrich09}, a fraction of stars may belong to the thick disc, to an older 
thin-disc component at lower metallicities \citep[see][]{Zhang25}, or even to the stellar halo. 
Among the most distant stars ($\gtrsim 3$ kpc, $\approx$30\% of the sample), some could belong to 
the tail of the bar or bulge component (results for a subsample with $\varpi > 0.4$ mas are shown 
in Appendix~\ref{app:alternative} and discussed in Section~\ref{sec:disc:cuts}). From a qualitative 
inspection of the sample kinematics, we estimate the contamination level to be $\lesssim$20--30\% 
(see Section~\ref{sec:systematics}). Taking this into account, we consider a binary fraction of 
$\gtrsim$0.7 to be sufficiently high to be consistent with an exclusively binary-formation origin 
for metal-rich RRLs, as predicted by \citet{BI24}.

\subsection{The RR Lyrae binary models} \label{sec:binmodel}

Stellar properties (masses and flux ratios) and binary parameters (orbital period and eccentricity) are drawn from detailed binary evolution simulations performed with MESA \citep[Modules for Experiments in Stellar Astrophysics;][]{Paxton11,Paxton13,Paxton15,Paxton18,Paxton19}, as presented in \citet{BI24} and originally introduced in \citet{Vos20}.

In these models, close low-mass binaries ($M<2\,M_\odot$, $P_\mathrm{orb}<700$ days) are predicted to produce stripped and partially stripped core–helium-burning stars in binaries with main-sequence companions. Among 2060 simulated systems, 22 fall within the instability strip, forming a population of metal-rich ($\mathrm{[Fe/H]}>-1$), intermediate-age thin-disc RRLs. 

After excluding a few cases where the companion outshines the RRL, we use this catalogue \citep{BI24cat}\footnote{Available on Zenodo at \url{https://doi.org/10.5281/zenodo.17398542}} to infer a multivariate posterior distribution over five parameters: orbital period ($P_\mathrm{orb}$), RRL mass ($M_\mathrm{RRL}$), companion mass ($M_\mathrm{comp}$), metallicity, and
$G$-band luminosity ratio ($f_\mathrm{lum} = 10^{0.4(G_\mathrm{RRL} - G_\mathrm{comp})}$), where the magnitudes are estimated from MESA and correspond to the time-averaged luminosities.
The joint distribution of these parameters is modelled using a Dirichlet Process Gaussian Mixture Model (DPGMM), implemented with the \texttt{Figaro} package \citep{Rinaldi24}\footnote{We use version~1.10.2 from the \texttt{PyPi} repository: \url{https://pypi.org/project/figaro/1.10.2/}.}. Unlike standard Gaussian Mixture Models, the DPGMM infers the number of components directly from the data, allowing a flexible reconstruction of complex, multi-modal distributions \citep{Teh10}. Using default settings, we generate $10^5$ DPGMM realisations, which are then used to sample binary parameters conditioned on the photometric metallicity of each RRL (see Appendix~\ref{app:DPGMM} for additional details on the inferred posterior distribution and sampling).

\subsection{Mock Gaia observations} \label{sec:gaiamock}

\subsubsection{The \texttt{gaiamock} module} \label{sec:gaiamockmod} 

To simulate Gaia astrometric observations, we use the forward-modelling framework \texttt{gaiamock} \citep{ElBadry24mock}, which generates synthetic Gaia epoch astrometry for Gaia DR3 and future releases by modelling the Gaia scanning law, astrometric uncertainties, and photocentre motion induced by unresolved binaries. The simulated epoch astrometry is fitted using the same hierarchy of astrometric models adopted in Gaia DR3 \citep{GaiaDR3binary}, including single-source solutions and non-single-source models of increasing complexity: accelerated (7-parameter), variable-acceleration (9-parameter), and full orbital (12-parameter) solutions.

For each fit, \texttt{gaiamock} returns best-fit parameters together with the standard Gaia diagnostics, including the goodness-of-fit statistic $F_2$ and the solution significance $s$, as defined in the Gaia DR3 non-single source pipeline \citep{GaiaDR3binary}. The module also provides the Unit Weight Error (UWE), defined as the square root of the reduced chi-squared of the astrometric fit. 
\rev{In practice, due to calibration errors, the Gaia UWE for well-behaved sources depends on the colour and magnitude of the source; hence its renormalised version, \ruwe, is often used instead \citep{Lindegren21}. Since in \texttt{gaiamock} UWE is not affected by calibration errors (apart from the variability-induced biases introduced in our analysis), it is directly comparable to \ruwe reported in the Gaia catalogue.}
Hereafter, we use \ruwe to refer to both the Gaia-reported value and to the corresponding UWE estimated with \texttt{gaiamock}. A system is flagged as detectable if any of the non-single source models is accepted following the Gaia DR3 astrometric cascade.

The standard implementation of \texttt{gaiamock} does not account for intrinsic stellar variability. Since RRLs exhibit large-amplitude photometric and colour variations, we employ a customised version of the code that includes variability-induced astrometric effects, such as heteroskedastic uncertainties, photocentre shifts driven by time-dependent flux ratios, and residual chromatic biases. 
In addition, we also include an update for the \ruwe estimate.

\subsubsection{\texttt{gaiamock} update:  heteroskedastic errors and variability-induced mover  effect and fit} \label{sec:vim}

To account for epoch-to-epoch variations in astrometric uncertainties and for the variability-induced mover (VIM) effect, we modify the generation of astrometric data for both single and binary sources. In this framework, astrometric errors (which depend on the $G$-band magnitude) and the flux ratio are no longer fixed at their mean values, but instead vary at each epoch following the intrinsic stellar variability.
In Gaia DR3, a binary astrometric model based on the VIM effect was tested \citep[VIMF model; see][]{GaiaDR3binary}. We have implemented this model in \texttt{gaiamock}. Appendix~\ref{app:vim} describes the details of the VIM effect and the VIMF model fitting procedure.

\subsubsection{\texttt{gaiamock} update: chromaticity effect} \label{sec:chrom}

Because the Gaia mirrors are not perfectly achromatic, stars of different colours are measured at slightly different apparent positions \citep{Lindegren22}. In the Gaia pipeline, this effect is calibrated using the source average colour \citep{Lindegren21}\footnote{The chromatic calibration in Gaia is not based directly on the observed colour, but on an effective pseudo-colour derived from the astrometric solution itself, which ensures self-consistency and avoids propagating photometric noise into the astrometry, see \citealt{Lindegren21} for additional details.}, so colour variations due to random noise are absorbed into the astrometric noise budget already modelled in \texttt{gaiamock}. As a result, no additional treatment is required for non-variable sources.

For variable stars with periodic colour changes, however, chromatic shifts cannot be treated as purely random noise. In this case, the astrometric displacement depends explicitly on the observation time, introducing correlated residuals that can affect the astrometric solution, the goodness-of-fit statistics, and ultimately the probability of a source being classified as a binary.

To model this effect, we modify the standard \texttt{gaiamock} along-scan (AL) astrometric predictions for both single and binary sources as
\begin{equation}
\begin{aligned}
\eta_\mathrm{var}(t) &=
\eta(t) \\
&\quad + f_\mathrm{chrom}
\Big[(G_\mathrm{BP}-G_\mathrm{RP})(t) - \langle G_\mathrm{BP}-G_\mathrm{RP}\rangle\Big] \\ 
&\quad +  \eta_\mathrm{noise}(G(t)),
\end{aligned}
\label{eq:etavar}
\end{equation}
where $\eta(t)$ is the standard AL position from \texttt{gaiamock}, and  $\eta_\mathrm{noise}$ are the (heteroskedastic) astrometric errors depending on the $G$-bad magnitude at specific time of observation (see Appendix \ref{app:vim}).
The chromatic term is assumed to scale linearly with the deviation of the instantaneous colour from its mean value, as predicted by the variability model. Its amplitude is set by the free parameter $f_\mathrm{chrom}$ (in units of mas mag$^{-1}$), representing the astrometric shift induced by a colour change of 1 mag.

Although the Gaia chromatic calibration is more complex, it is ultimately based on combinations of linear terms \citep{Lindegren18}. Equation~\ref{eq:etavar} therefore provides a simplified but conceptually consistent representation of residual chromatic variability effects, while retaining a simple, easily interpretable form. The strength of the effect is controlled by $f_\mathrm{chrom}$, which we vary to explore its impact on the inferred astrometric solutions.

\subsubsection{\texttt{gaiamock} update: \ruwe estimate} \label{sec:ruwe}

The version of \texttt{gaiamock} described by \citet{ElBadry24mock} robustly reproduces the observed \ruwe\ values of binaries whose astrometric residuals are dominated by orbital motion \citep[e.g.][]{ElBadry25}. However, it does not reliably predict the distribution of \ruwe\ for sources with well-behaved single-star fits (i.e., \ruwe\ $\approx 1$). In particular, observed sources have a broader \ruwe\ distribution than predicted, and the median \ruwe\ varies across the sky, showing both imprints of the Gaia scanning law and variations in local source density \citep[e.g.][]{Castro24}. We make two modifications to \texttt{gaiamock} to better reproduce the observed \ruwe\ distribution in the low-\ruwe\ regime.

In the original version of \texttt{gaiamock}, the epoch astrometric measurements were treated at the level of a single field-of-view (FOV) transit, effectively averaging the $\approx 8$ usable CCD measurements collected during each transit and reducing the uncertainty by a factor $\sqrt{8}$. While this preserves the nominal signal-to-noise ratio, it modifies the effective number of measurements entering the fit and therefore biases the reduced $\chi^2$ and the derived \ruwe.

In the updated implementation, we explicitly model the individual CCD observations when computing \ruwe\ and the associated goodness-of-fit statistics. This ensures that the resulting \ruwe\ values are more closely comparable to those reported in the Gaia catalogue, particularly for well-behaved single sources.

In addition, we introduce a rescaling of the astrometric uncertainties, $\sigma_\eta$, to reproduce the observed sky-dependent variations of \ruwe\ in Gaia DR3:
\begin{equation}
\sigma_\eta(G,\alpha,\delta) =
\sigma_{\eta,\rm CCD}(G)
+
\left[\ruwe_{\rm med}(\alpha,\delta)-1\right]
\sigma_{\eta,\rm sky},
\end{equation}
where $\sigma_{\eta,\rm CCD}$ is the along-scan uncertainty per CCD transit (as in the original \texttt{gaiamock} implementation), $\ruwe_{\rm med}$ is the Gaia DR3 median \ruwe\ as a function of sky position $(\alpha,\delta)$ for sources with $11 < G< 13$, and $\sigma_{\eta,\rm sky}=0.125\,\mathrm{mas}$ corresponds to the sky-averaged value of $\sigma_\eta$ at $G=12$. We obtain $\ruwe_{\rm med}(\alpha,\delta)$ from a tabulated healpix map of level 4 (3072 regions).

 We verified that this procedure, which treats variations in \ruwe\ across the sky as a result of variations in the Gaia astrometric precision, reliably reproduces the observed sky-dependence of \ruwe\ for both bright and faint sources. The root cause of this sky dependence is not yet well understood, but our experiments demonstrate that it can be effectively modelled as a position-dependent rescaling of the astrometric uncertainties. 

Although this procedure provides a more robust estimate of \ruwe, it is computationally more expensive and increases the cost of the binary detectability analysis by about a factor of ten.
For this reason, when assessing detectability (see Sections~\ref{sec:mock} and \ref{sec:detectability_definition}), we use the original \texttt{gaiamock} version including the updates for the chromaticity (Section \ref{sec:chrom}) and VIM (Section \ref{sec:vim}) effects.
The original \ruwe estimate has also been used in the companion paper by \cite{Pranav26} (but without the two aforementioned effects), and it has been shown to provide a realistic match to the observed properties of binary sources in Gaia \citep[see e.g.][]{ElBadry24,Muller25}.

\subsection{Mock generation and explored models} \label{sec:mock}

\begin{table}
\centering
\caption{Summary of the explored models and their adopted parameters.}
\begin{tabular}{lc}
\hline \hline
Label & Parameters \\
\hline

R      & Reference model \\

V065   & $f_\mathrm{chrom}=0.65$ \\

V03    & $f_\mathrm{chrom}=0.30$ \\

P03    & $f_\mathrm{period}=0.3$ \\

P2     & $f_\mathrm{period}=2.0$ \\

E5     & $e=0.5$ \\

V065P2 & $f_\mathrm{chrom}=0.65$, $f_\mathrm{period}=2.0$ \\

NV     & not variable \\

\hline
\multicolumn{2}{c}{\textit{Only single stars}} \\
\hline

SR     & Reference model \\

SV065  & $f_\mathrm{chrom}=0.65$ \\

SV03   & $f_\mathrm{chrom}=0.30$ \\

SNV    & not variable \\
\hline
\end{tabular}
\begin{tablenotes}
\small
\item Notes. The reference model is described in Section~\ref{sec:mock}. The parameter $f_\mathrm{chrom}$ controls the chromaticity-induced astrometric bias (Equation\ref{eq:etavar}), $f_\mathrm{period}$ rescales the orbital periods predicted by the binary-formation model, and $e$ denotes the orbital eccentricity. The label not variable indicates that no variability-induced astrometric bias is included (Section \ref{sec:gaiamock}).
\end{tablenotes}
\label{tab:models}
\end{table}

To estimate the detectability of each RRL in our sample under the assumption that it resides in a binary system, we perform Monte Carlo simulations.

For each star, we generate 1000 realisations of the observed Gaia DR3 properties (sky position, proper motions, parallax, photometric metallicity, mean magnitude, and colour), drawing from multivariate Gaussian distributions defined by the catalogue uncertainties and covariances. Parallaxes are corrected for the Gaia DR3 global zero-point offset, adopting $\varpi_\mathrm{zp}=-0.033$ mas \citep{Garofalo22}.

Each realisation is combined with binary properties drawn from the \citet{BI24} formation model (Section~\ref{sec:binmodel}). Binary parameters (component masses, $G$-band flux ratio, and orbital period) are sampled from metallicity-conditioned probability distributions inferred from the model (Appendix \ref{app:DPGMM}). The orbital configuration is completed by drawing the inclination from an isotropic distribution, $p(i)\propto\sin i$, and by drawing the remaining angular parameters from uniform distributions: the longitude of the ascending node $\Omega \sim \mathcal{U}(0,2\pi)$ and the argument of periastron $\omega \sim \mathcal{U}(0,2\pi)$. The time of periastron passage is drawn from a uniform distribution between 0 and the orbital period. All the models in \citet{BI24} assume circular orbits, which is a limitation of the model, while the real systems may well be eccentric.

To account for model variation systematics, we produce 8 different samples based on the different models whose labels and properties are summarised in Table \ref{tab:models}. 

The reference model (R) adopts the original \citet{BI24} predictions and includes variability-induced astrometric effects (Appendix \ref{app:vim}) but no chromaticity bias ($f_\mathrm{chrom}=0$).
Models V03 and V065 include chromaticity effects with $f_\mathrm{chrom}=0.3$ and $0.65$, corresponding to typical chromatic shifts of $\sim0.1$ and $\sim0.3$ mas for RRL colour variations of 0.2–0.4 mag.
To probe variations in the orbital properties, we also test a model in which we scale the periods sample from the model by a constant factor $f_\mathrm{period}$, testing shorter ($f_\mathrm{period}=0.3$, P03) and longer ($f_\mathrm{period}=2$, P2)  periods. We also test a model with high eccentricity ($e=0.5$, E5). 
Periods of 500–1000 days and eccentricities up to $e\sim0.5$ are commonly observed in post-mass-transfer binaries \citep{Moltzer25}, while longer periods of 2000–4000 days have been reported for RRL binary candidates \citep{Hajdu21}.
We further consider a combined model that includes both chromaticity bias and longer periods (V065P2) and a model that neglects variability effects altogether (NV).

We also produce four additional single-star models (SR, SV03, SV065, and SNV), which assume no binarity but include the same variability and chromaticity prescriptions as their binary counterparts.

\subsection{Bayesian analysis}

\subsubsection{Detectability} \label{sec:detectability_definition}

For a given model (Table \ref{tab:models}) and for each star in the sample, we process the 1000 realizations using the updated \texttt{gaiamock} module (Section~\ref{sec:gaiamock}). The variability model is based on the $G$-band and $(G_\mathrm{BP}-G_\mathrm{RP})$ light curves as provided by the Gaia DR3 catalogue \citep{Clementini23}. In practice, we reconstruct the light curves in the three Gaia photometric bands ($G$, $G_\mathrm{BP}$, and $G_\mathrm{RP}$) using the harmonic decomposition parameters reported in the Gaia DR3 RRL catalogue. Given these models, the \texttt{gaiamock} tools retrieve the model-predicted magnitude at each transit. As in \cite{ElBadry24}, we randomly reject FOV transits with a 10\% probability to account for the fact that in Gaia DR3, about 10\% of the transits are not used in the astrometric pipeline.

Each realisation is fitted with all the astrometric models explored in Gaia DR3. From the single source solution, we compute the predicted \ruwe. To estimate the detectability of each source in Gaia, we use the fitted solutions to reproduce the Gaia DR3 astrometric ``cascade'' of model selection as described in \citealt{ElBadry24}. 
The only modification to the standard cascade is the inclusion of the VIMF fit (Appendix \ref{app:vim}). The VIMF fit is attempted only when the orbital solution is not accepted, and it is accepted only if $F_2 < 25$, $s > 20$ (see Section \ref{sec:gaiamockmod}), and $\varpi/\sigma_\varpi > 30$ (parallax signal-to-noise). If the VIMF model is not accepted (or if the 9-parameter or 7-parameter solutions are accepted but with $s<20$), we flag the current realisation as not detectable as an astrometric binary in Gaia.

The detection process is modelled as a Bernoulli experiment, following a Binomial distribution with $N_\mathrm{r}$ trials and an unknown success probability $\mathcal{O}$.
Despite the classical frequentist connotation of this term, in our Bayesian framework   
$\mathcal{O}$ is treated as a stochastic variable; we refer to it hereafter as the detectability. 
Adopting Bayes’ theorem and assuming a Beta prior on $\mathcal{O}$, $P(\mathcal{O})\sim\mathcal{B}(1,1)$\footnote{Equivalent to a uniform distribution on $[0,1]$.}, the posterior distribution for the detectability is also a Beta distribution \citep{Gelman2014}:
\begin{equation}
P(\mathcal{O}|N_\mathrm{b},N_\mathrm{r}) \sim \mathcal{B}(1+N_\mathrm{b},1+N_\mathrm{r}-N_\mathrm{b}),
\label{eq:detectability}
\end{equation}
where $N_\mathrm{b}$ is the number of realisations yielding a positive binary detection (i.e. accepted by at least one of the tested astrometric binary models), and $N_\mathrm{r}=1000$ is the total number of realisations.

\subsubsection{Binary fraction posterior} \label{sec:fbinmet}

\begin{figure*}
    \centering
\includegraphics[width=\linewidth]{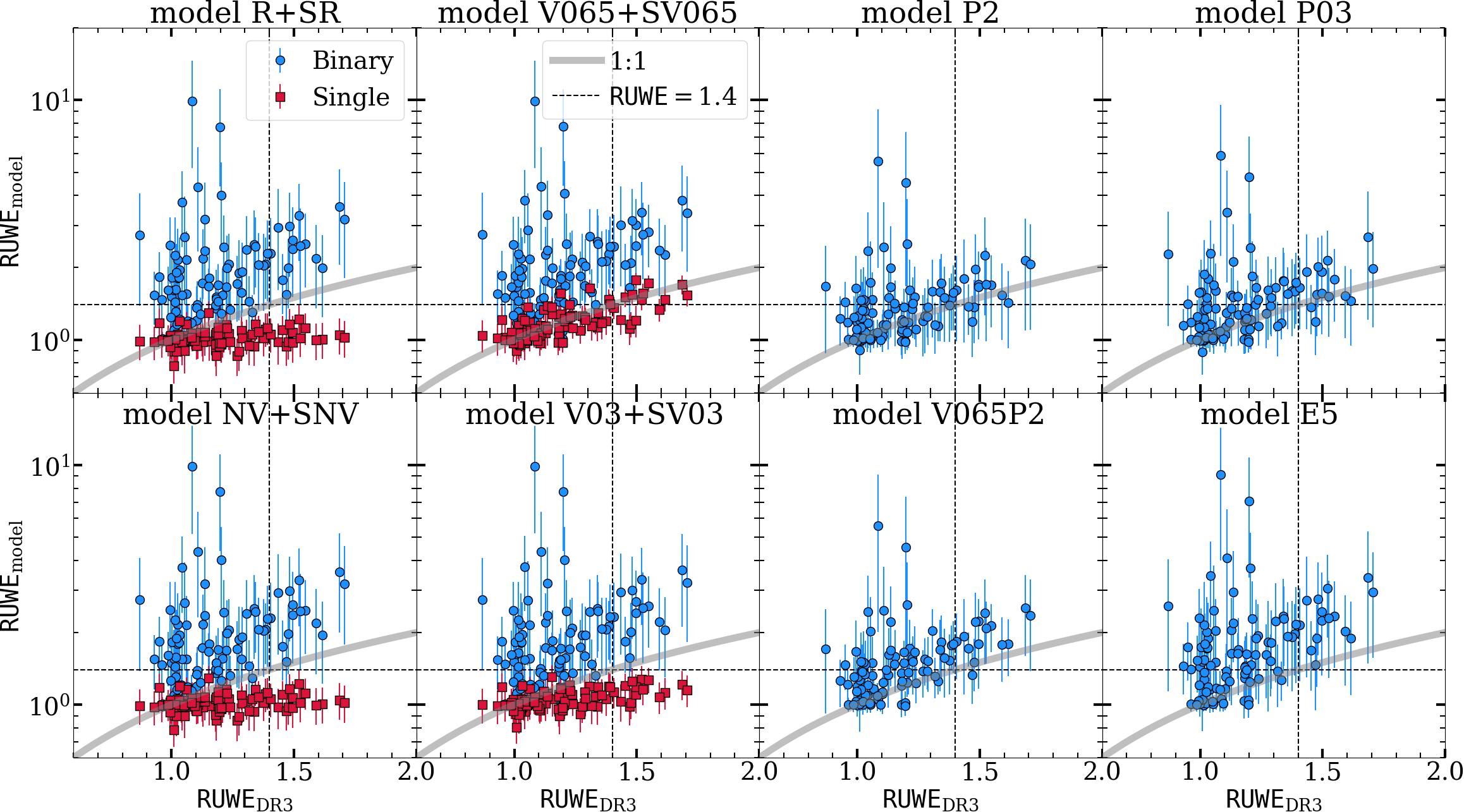}
    \caption{Comparison between model-predicted \ruwe\ ($\ruwe_\mathrm{model}$) and Gaia DR3 values ($\ruwe_\mathrm{DR3}$) for the RRLs in the analysed sample (see panel titles and Table~\ref{tab:models}). The four left-hand panels show both the binary model (blue circles) and the single-star model (red squares); the remaining panels show the binary model only. Markers indicate the mean of 1000 realisations, with error bars corresponding to 2 times the standard deviation ($\approx95\%$ confidence interval). The thick grey line marks $\ruwe_\mathrm{model}=\ruwe_\mathrm{DR3}$, and the black dashed line indicates $\ruwe=1.4$.}
    \label{fig:ruwehist}
\end{figure*}

For a star drawn from a population with binary fraction $f_\mathrm{bin}$, the probability of detecting it as a binary is $f_\mathrm{bin}\mathcal{O}$, where $\mathcal{O}$ is the detectability of that source (we neglect false positives).
In our case, we treat both the detectability and the binary fraction as stochastic variables. Detectability follows the distribution defined in Equation~\ref{eq:detectability} and is conditioned on the positive detections obtained from the \texttt{gaiamock} analysis (Section~\ref{sec:mock}). The binary fraction is instead conditioned on the number of binaries detected by Gaia, and constraining its posterior distribution is one of the main goals of this work.

For a sample of $N$ stars, the likelihood of detecting a specific set $N_\mathrm{D}$ binaries (of which we know the identity, i.e. we condition on which sources are detected rather than on the total count alone) is obtained by marginalising over the detectability of each source:

\begin{equation}
P(f_\mathrm{bin},\vec{d})=  \mathcal{L}(f_\mathrm{bin}) =
\prod^{N}_{i}   \int^1_0 (f_\mathrm{bin}\,\mathcal{O})^{d_i} (1-f_\mathrm{bin}\,\mathcal{O})^{(1-d_i)} 
P_i(\mathcal{O}) d \mathcal{O},
\label{eq:lkl2}
\end{equation}
where $P_i(\mathcal{O})$ is the posterior of the detectability for the $i$-th star (Equation \ref{eq:detectability}), and $\vec{d}$ is a vector with $d_i=1$ for systems detected as binary and $d_i=0$ for all the others.

Since $f_\mathrm{bin}$ and $\mathcal{O}$ are independent, the integrals reduce to the mean detectability $\langle \mathcal{O}\rangle_i$ of each source. In our case, no binaries are detected ($N_\mathrm{D}=0$), so the likelihood simplifies to
\begin{equation}
\mathcal{L}(f_\mathrm{bin}) = \prod_{i=1}^{N} \left(1 - f_\mathrm{bin}\langle \mathcal{O}\rangle_i\right),
\label{eq:lkl}
\end{equation}
where $\langle \mathcal{O}\rangle_i = (1+N_{\mathrm{b},i})/(2+N_{\mathrm{r},i})$ is the mean of the Beta posterior inferred from the Monte Carlo simulations (Equation~\ref{eq:detectability}).

Assuming a uniform prior on $f_\mathrm{bin} \sim \mathcal{U}(0,1)$, the posterior distribution of the binary fraction is fully determined by Equation~\ref{eq:lkl}. Because the likelihood decreases monotonically with $f_\mathrm{bin}$, the posterior mode is trivially at $f_\mathrm{bin}=0$. However, in the absence of detections, this point estimate is not informative. We therefore focus on upper limits and credible intervals, which quantify the maximum binary fraction compatible with no binary detection in the data. Our goal is not to measure the true binary fraction, but to assess whether a high fraction of binaries could remain hidden in the sample.

Motivated by the predictions of \citet{BI24} and by observational evidence, we investigate how the maximum allowed binary fraction depends on metallicity using two complementary strategies. 
First, we divide the sample into two metallicity bins split at [Fe/H]=-0.3 and infer the posterior distribution of $f_\mathrm{bin}$ independently in each bin using Equation~\ref{eq:lkl}. This is achieved through Monte Carlo sampling.
For each realisation, we draw each star's metallicity from a Gaussian centred on its photometric metallicity estimate with width equal to its measurement uncertainty, and draw the binary detectability from the distribution given by Equation \ref{eq:detectability}. Only stars whose drawn metallicity falls within the target range are retained.
Repeating this procedure $1000$ times allows us to reconstruct the posterior distribution of $f_\mathrm{bin}$ and derive the credible intervals reported in Table \ref{tab:models_bfrac}

In parallel, we allow the binary fraction to vary continuously with metallicity by modelling $f_\mathrm{bin}$ as an unknown smooth function of metallicity. This dependence is described using a Sparse Gaussian Process (GP) as a latent function, mapped onto the physical interval $[0,1]$ through a logistic transformation,
\begin{equation}
f_\mathrm{bin}(x) =
\frac{1}{1+\exp\left[-k\left(f^\ast(x)-f_0\right)\right]},
\label{eq:fbin}
\end{equation} where
\begin{equation}
x = \frac{\left( \mathrm{[Fe/H]} - \langle \mathrm{[Fe/H]} \rangle \right)}{\sigma_{\mathrm{[Fe/H]}}}
\label{eq:xfeh}
\end{equation}
is the rescaled metallicity with respect to the sample metallicity mean ($\langle \mathrm{[Fe/H]} \rangle$) and standard deviation ($\sigma_{\mathrm{[Fe/H]}}$)
The latent function $f^\ast(x)$ is drawn from a GP with a squared-exponential kernel, characterised by the amplitude $a_\mathrm{GP}$ and correlation length $\ell_\mathrm{GP}$. The parameters $a_\mathrm{GP}$, $\ell_\mathrm{GP}$, $k$, and $f_0$ are treated as free parameters and marginalised over.

To reduce computational cost and mitigate overfitting, we adopt a Sparse GP approximation with 15 inducing points using the Fully Independent Training Conditional (FITC) method \citep{Candela2005}. 
The 15 inducing points are uniformly spaced in the metallicity range covered by the data. 
The full Bayesian model, including marginalisation over metallicity uncertainties and detectability, is implemented and sampled using PyMC \citep{PYMC}\footnote{Version 5.24.1}.
All the details are provided in Appendix~\ref{app:fitc} and summarised in Table \ref{tab:fitc}.

Posterior sampling is performed using the No-U-Turn Sampler (NUTS), an adaptive Hamiltonian Monte Carlo (HMC) algorithm \citep{HoffmanGelman2014NUTS}, running 4 chains with 2000 tuning (burn-in) steps and 5000 posterior draws per chain. We assess MCMC convergence using the Gelman--Rubin statistic $\hat{R}$, requiring $\hat{R}<1.01$ for all sampled parameters.

Within this framework,  the posterior on the number of predicted detections, P($N_\mathrm{D}$), is naturally obtained as the sum of independent Bernoulli trials following a Poisson-Binomial distribution whose mean and variance, under the assumption of constant binary fraction, are reported in Equations \ref{eq:Ndet} and \ref{eq:Nvariance}.

\section{Results} \label{sec:results}

\subsection{\ruwe} \label{sec:ruwe}

Figure~\ref{fig:ruwehist} compares the observed \ruwe\ distribution of our RRL sample with the model predictions. All binary models systematically predict higher \ruwe\ values, particularly those adopting the fiducial period distributions of \cite{BI24} (models R, NV, V065, and V03), for which the mean \ruwe\ offset is $\approx 0.8$–0.9.
However, when accounting for the model uncertainties (represented by $\sigma$, estimated as the standard deviation of the \ruwe\ distribution across realisations), a substantial fraction of systems remains statistically consistent with the observed \ruwe: $\approx 70$–80\% within $3\sigma$ ($\approx 99\%$ confidence interval), and $\approx 30$–50\% within $2\sigma$ ($\approx 95\%$ confidence interval).
The models R and NV show almost equivalent results, suggesting that the VIM effect has a negligible impact, while model V065 shows only a slight increase in \ruwe, indicating that for binaries the \ruwe\ values are primarily driven by binarity rather than variability.

The models with longer (P2, V065P2) or shorter (P03) orbital periods provide an overall better match to the observed \ruwe. The mean offset reduces to $\approx -0.17$ for P03, to $\approx 0.25$ for P2, and $\approx 0.37$ for V065P2. In all these cases, a larger fraction ($\approx 90$–95\%) of systems is consistent with the observed \ruwe\ within $3\sigma$, and in the case of P2, about half of the sample is consistent within $1\sigma$.

The case of higher eccentricity (E5) lies between the two subgroups, with an average offset ($\approx 0.7$) slightly lower than that of the models with fiducial orbital parameters, but with a higher fraction of systems ($\approx 92\%$) consistent within $3\sigma$.

Figure~\ref{fig:ruwehist} shows that in all models there is a subgroup of objects ($\approx 10\%$) with very high predicted \ruwe\ ($\gtrsim 3$), with a couple of systems (Gaia DR3 2857456211775108480, also known as SW And, and Gaia DR3 6005656897473385600, also known as FW Lup) reaching extreme values ($>8$). These correspond to systems with high parallax ($\varpi \gtrsim 0.7$) and high parallax signal-to-noise ($\gtrsim 40$), representing the closest sources ($\lesssim 2$ kpc), with SW And and FW Lup located within 500 pc.
For these cases, especially the latter two, the models predict that orbital periods in the range 300–3000 days and with eccentricity $< 0.5$ should already have manifested as a high \ruwe, exceeding the observed values. It is worth noting that  both stars are  somewhat peculiar. SW And has some mass estimates as low as $0.26\,$M$_\odot$ \citep{Barcza14},  which is not consistent with the binary-formation model of \cite{BI24}, and FW Lup has been noted as a particularly low-amplitude pulsator \citep{Nemec24}.

Concerning the single-star models, Figure~\ref{fig:ruwehist} clearly shows that models not including the chromaticity effect (SR, SNV), or assuming only a mild impact (SV03), systematically produce lower \ruwe\ values. Higher \ruwe\ values are often interpreted as signatures of unresolved binarity. However, \citet{Belokurov20} showed that the \ruwe\ distribution of RRLs correlates with mean magnitude and variability amplitude, indicating that variability-induced biases can significantly affect the Gaia astrometric solution.
Indeed, the SV065 model provides the best match to the data, with an almost zero offset ($\approx 0.01$) and a dispersion broadly consistent with the estimated uncertainties. This is a remarkable result, especially in light of the simplified modelling of the chromaticity bias (Section \ref{sec:chrom}).
It also offers a quantitative explanation for the empirical trends reported by \citet{Belokurov20}, with chromaticity effects being more important at bright magnitudes, where they dominate the astrometric noise budget, and becoming negligible for fainter stars ($G > 16$).

Considering the \ruwe\ comparison, the single-star model is the one that best reproduces the observations, while the binary models tend to systematically over-predict \ruwe\ unless longer or shorter orbital periods relative to the \cite{BI24} models are adopted. However, given the uncertainties, no binary model can be ruled out at high significance ($>3\sigma$) for $\approx 80$–90\% of the sources, although a few nearby metal-rich RRLs (e.g. SW And and FW Lup) already indicate a strong signature of non-binarity due to their low \ruwe ($<1.2$) when compared with the model predictions.

\subsection{Detectability} \label{sec:detectability}

\begin{figure*}
    \centering
    \includegraphics[width=\linewidth]{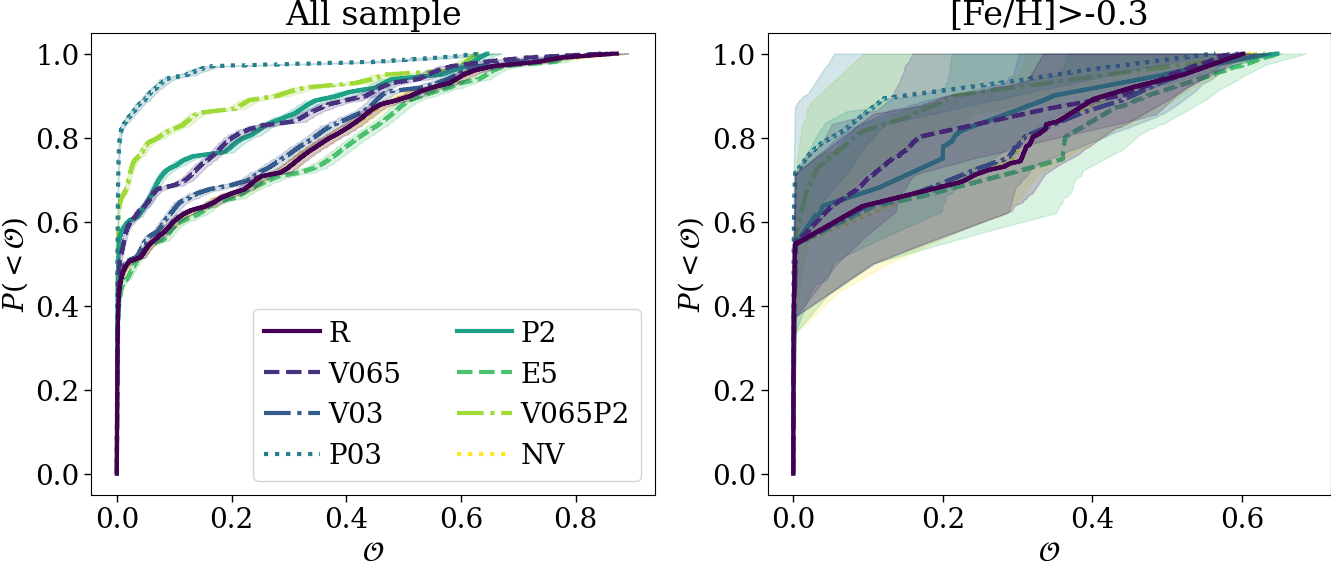}
     \caption{Median (lines) and 95\% credible intervals (shaded regions) of the cumulative distribution function (CDF) of the Gaia DR3 detectability, $\mathcal{O}$, which defines the probability of receiving an astrometric binary detection (see Section~\ref{sec:detectability_definition}). The left panel includes all RRLs in the analysed sample, while the right panel additionally applies a metallicity cut $\mathrm{[Fe/H]}>-0.3$. Colours and line styles correspond to the models listed in Table~\ref{tab:models}. Posteriors are obtained by drawing $10^4$ samples per star from Equation~\ref{eq:detectability} and sampling the photometric metallicities assuming Gaussian uncertainties.}
    \label{fig:detdr3}
\end{figure*}

\begin{table*}
\centering
\small
\setlength{\tabcolsep}{5pt}
\caption{Summary of posterior constraints on $f_\mathrm{bin}$ and predicted detections}
\begin{tabular}{l|ccc c|ccc c|ccc c}
\hline
 & \multicolumn{4}{c|}{Whole sample}
 & \multicolumn{4}{c|}{[Fe/H]$<-0.3$}
 & \multicolumn{4}{c}{[Fe/H]$>-0.3$} \\
\hline
 & \multicolumn{3}{c}{$q_{f_\mathrm{bin}}$} & $N_\mathrm{min}(f_\mathrm{bin}=0.7)$
 & \multicolumn{3}{c}{$q_{f_\mathrm{bin}}$} & $N_\mathrm{min}(f_\mathrm{bin}=0.7)$
 & \multicolumn{3}{c}{$q_{f_\mathrm{bin}}$} & $N_\mathrm{min}(f_\mathrm{bin}=1.0)$ \\

Label 
 & 0.68 & 0.95 & 0.99 & 
 & 0.68 & 0.95 & 0.99 & 
 & 0.68 & 0.95 & 0.99 &  \\
 
\hline

R      & 0.07 & 0.17 & 0.31 & 5 
       & 0.10 & 0.25 & 0.45 & 2 
       & 0.19 & 0.46 & 0.77 & 0 \\

V065   & 0.09 & 0.24 & 0.43 & 2
       & 0.14 & 0.34 & 0.61 & 0 
       & 0.27 & 0.61 & 0.90 & 0 \\

V03    & 0.07 & 0.18 & 0.34  & 4
       & 0.11 & 0.27  & 0.48 & 2 
       & 0.21 &  0.49 & 0.81  & 0  \\

P03    & 0.33 & 0.73 & 0.97   & 0 
       & 0.44  & 0.84   & 0.99   & 0 
       & 0.52 & 0.89  & 0.99   & 0  \\

P2     & 0.10 & 0.26  & 0.48  & 2  
       & 0.18 & 0.39  & 0.69  & 0  
       & 0.27 & 0.62  & 0.91 & 0  \\

E5     & 0.06 & 0.16  & 0.29  & 6  
       & 0.09 & 0.23 & 0.42 & 2  
       & 0.18 & 0.43  & 0.73  & 0  \\

V065P2 & 0.15  & 0.36  & 0.65 &  0
       &  0.21  & 0.51  & 0.83 & 0  
       & 0.38 & 0.77   & 0.97  & 0   \\

NV     & 0.07 & 0.17 & 0.31 & 5 
       & 0.10 & 0.24  & 0.44  & 2 
       & 0.19 & 0.46  & 0.78  & 0  \\

\hline
\end{tabular}
\begin{tablenotes}
\small
\item Notes. For all tested models, we list the posterior quantiles ($q_{f_\mathrm{bin}}$) of the binary fraction $f_\mathrm{bin}$ (Equation~\ref{eq:lkl}) assuming zero detections in Gaia DR3, and the corresponding minimum expected number of detections, $N_\mathrm{min}$, evaluated at the $f_\mathrm{bin}$ value given in the table header. Uncertainties arising from random bootstrap resampling of the detectability (Equation \ref{eq:detectability}) and of the photometric metallicity are negligible for the full sample ($<0.1\%$), while they reach levels of $\simeq 5\%$ and $\simeq 10\%$ for the lower-metallicity ($\mathrm{[Fe/H]} < -0.3$) and higher-metallicity ($\mathrm{[Fe/H]} > -0.3$) subsamples, respectively. The values of $N_\mathrm{min}$ are estimated from the 0.15th percentile of the posterior distribution of the expected number of detections (Equations \ref{eq:Ndet} and~\ref{eq:Nvariance}), assuming an underlying population with $f_\mathrm{bin}=0.7$ for the full sample and the lower-metallicity subsample, and $f_\mathrm{bin}=1.0$ for the higher-metallicity sample. See Table~\ref{tab:models} for details of the different models.
\end{tablenotes}
\label{tab:models_bfrac}
\end{table*}

\begin{figure*}
    \centering
    \includegraphics[width=\linewidth]{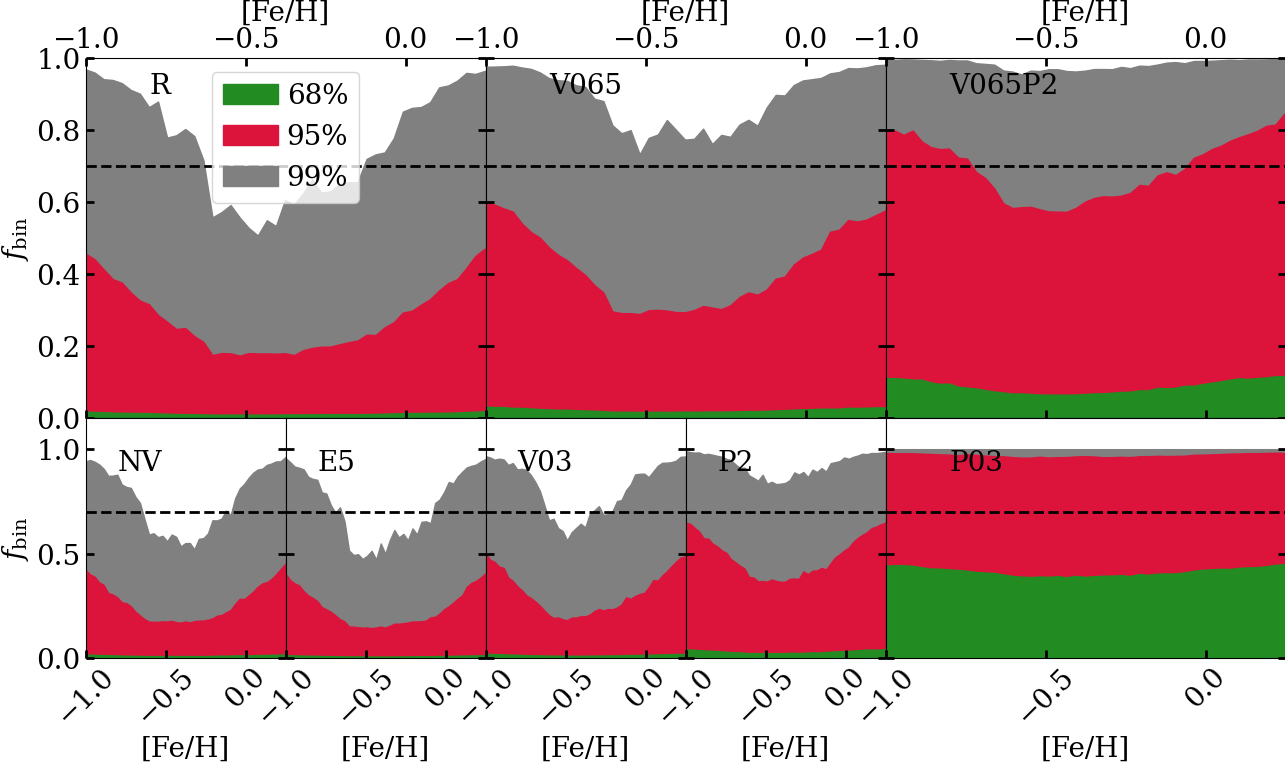}
    \caption{Binary-fraction posterior (Equation~\ref{eq:lkl}) as a function of metallicity, inferred using the metallicity-dependent Gaussian Process model (Equation~\ref{eq:fbin}). Shaded regions show 68\% (green), 95\% (red), and 99.5\% (gray) credible intervals. Each panel corresponds to a different model, as indicated in the panel labels (see Table~\ref{tab:models}).
    The black dashed line marks $f_\mathrm{bin}=0.7$, the binary fraction consistent with the binary formation model for metal-rich RRLs, assuming a conservative 30\% contamination (non metal-rich RRLs in the thin disc and non-RRL contaminants; see Section~\ref{sec:sample}).}
    \label{fig:fbin}
\end{figure*}

Figure~\ref{fig:detdr3} shows the cumulative Gaia DR3 detectability for all tested models. The reference (R), eccentric (E5), and non-variable (NV) models yield the highest detectability, with $\approx 20\%$ of the sample exceeding $\mathcal{O} > 0.5$.
The model in which we assume a mild impact of chromaticity bias (V03) produces only a slight decrease in detectability with respect to the reference model.
In contrast, models with shorter orbital periods (P03), or longer orbital periods combined with chromaticity bias (V065P2), have detectabilities below 0.5 for more than 95\% of the sources. Among these, model P03 is by far the most pessimistic, with $\gtrsim 90\%$ of the distribution within $\mathcal{O} = 0.1$.
The V065 and P2 models have similar detectability distributions, lying between the two extremes, with  $\lesssim 5$–10\% of sources reaching $\mathcal{O} > 0.5$.
Restricting the analysis to the most metal-rich subsample with [Fe/H]$>-0.3$ (right panel of Figure~\ref{fig:detdr3}) suppresses the high-detectability tail, with the maximum detectability reduced to $\mathcal{O} \approx 0.6$ across all models.

Overall, all models predict that at least $\approx 70\%$ of the sample has $\mathcal{O} < 0.5$, making binary detection challenging. Nonetheless, a few sources, primarily among the closest ones, have a high detectability. For example, SW And and FW Lup, the two nearby sources already discussed in Section~\ref{sec:ruwe}, have detectability ranging from $\mathcal{O} \approx 0.5$–0.6 in the most pessimistic cases to $\mathcal{O} \gtrsim 0.8$ in the reference model. In the reference model, a total of three stars have $\mathcal{O} \gtrsim 0.7$ (SW And, FW Lup, and HU Cas).
Consistently, Table~\ref{tab:models_bfrac} shows that, assuming a high-binary fraction ($f_\mathrm{bin}=0.7$), a null detection is allowed at the 99\% confidence level only for the models P03 and V065P2. 
However, considering the smaller subsample of RRLs with the highest metallicity ([Fe/H]$>-0.3$,  $\sim$ 10 objects), all the models are consistent with 0 detections within 3$\sigma$.

Considering the detectability channels, for most models, the large majority of binaries ($\gtrsim 98\%$) are detected with a full orbital solution. A small fraction ($\lesssim 0.4\%$) are recovered with 9-parameter solutions, while 7-parameter solutions are negligible and VIMF detections are absent, except for a few rare cases in model P2 ($\lesssim 0.01\%$).
The main exceptions are the long-period models (P2 and V065P2), for which the fraction of full orbital solutions decreases to $\approx 30\%$ for P2 and $\approx 40\%$ for V065P2, while 7-parameter solutions become significant, accounting for $\gtrsim 60\%$ of the detections. In the short-period model (P03), instead, all detections correspond to full orbital solutions.
The absence of VIMF solutions (see Appendix~\ref{app:vim}) highlights that this detectability channel is essentially negligible for RRLs in binaries with periods shorter than $\sim 3000$ days.
We find no significant trends with variability. Variability-induced biases generally make the acceptance of astrometric binary solutions more difficult, but they do not preferentially favour or suppress any specific solution type.

\subsection{Binary fraction} \label{sec:fbin}

Table~\ref{tab:models_bfrac} shows that the posterior constraints on the binary fraction closely track the detectability results discussed in Section~\ref{sec:detectability}.
Considering the full sample, models with fiducial orbital periods and no or mild chromatic bias (R, NV, E5, V03) yield an upper limit on the binary fraction of $\approx 0.3$. A stronger chromaticity bias (V065) or longer orbital periods (P2) pushes this limit to $\approx 0.5$, while in the cases of V065P2 and P03 the constraints become substantially weaker, remaining consistent with binary fractions $\gtrapprox 0.7$, in line with the prediction of an exclusive production of metal-rich RRLs through binary formation in the sample which may be up to $30\%$ contaminated by the classical metal-poor RRLs. When restricting the analysis to the most metal-rich subsample ($\mathrm{[Fe/H]} > -0.3$), the constraints weaken further, partly due to smaller number statistics. In this regime, for all models, a high binary fraction (at least $\approx 0.7$) cannot be excluded at high statistical significance.

Figure~\ref{fig:fbin} presents the results obtained by allowing the binary fraction to vary continuously with metallicity using a fully Bayesian framework that also accounts for the uncertainties in detectability and metallicity, as well as their correlation (Section~\ref{sec:fbinmet}). 
Most models show a U-shaped posterior, with an increase in the allowed binary fraction at both the high- and low-metallicity ends.
\rev{Because the posterior strongly depends on the number of sources (Equation~\ref{eq:lkl}), this trend is largely driven by low-number statistics. Indeed, the bulk (95\%) of the RRLs in the sample have nominal photometric metallicities between $-0.6$ and $-0.2$ (Figure~\ref{fig:rrlsample}). However, given the large photometric uncertainties, the sampled metallicity range extends well beyond these nominal values: 95\% of the metallicities drawn from the individual error distributions fall between $-1.2$ and $0.35$. Thus, the tails of the metallicity distribution are still explored, albeit with considerably less statistical constraining power.}
Therefore, while a genuine increase in the binary fraction as a function of metallicity cannot be excluded within the uncertainties, the metallicity dependence of the posteriors shown in Figure~\ref{fig:fbin} should not be over-interpreted.
The model with the strongest reduction in detectability (P03) shows a flat metallicity trend, as the low detectability dominates over the effect of sample size.

The models in which chromaticity bias is not accounted for, or is assumed to have a low impact (V03), show that in the metallicity range with the highest source density a binary fraction higher than 0.6 is excluded at high significance.

In models with stronger chromaticity bias and/or shorter or longer orbital periods (V065, V065P2, P2, P03), a high binary fraction ($\approx 0.8$), consistent with the binary-formation scenario for metal-rich RRLs, cannot be excluded at the $3\sigma$ level.
This is because, in these models, the predicted orbital periods are either significantly shorter or significantly longer than the Gaia DR3 observing baseline ($\sim 1000$ days), preventing a clear detection of binary-induced astrometric signatures. In addition, strong chromaticity bias produces higher residuals, preventing the acceptance of binary astrometric solutions.
The most extreme model (P03) remains consistent with binary fractions up to $\approx 0.9$ within $2\sigma$, and $\approx 0.4$ at $1\sigma$.

\section{Discussion} \label{sec:discussion}

\subsection{Results interpretation} \label{sec:interpreation}

Although a high \ruwe\ value ($>1.4$) is a necessary condition for entering the Gaia DR3 non-single source pipeline, the two analyses presented here, the comparison of \ruwe\ (Figure\ref{fig:ruwehist}) and the inference of detectability and binary fraction (Figure\ref{fig:fbin}), probe complementary aspects.
Inclusion in the Gaia DR3 non-single source catalogue requires passing a stringent sequence of quality filters designed to suppress spurious solutions \citep{GaiaDR3binary}. Consequently, genuine binaries with elevated \ruwe\ values may still remain undetected.
Indeed, our results show that even when binarity inflates \ruwe, a significant fraction of systems can evade detection due to additional astrometric biases, most notably those induced by stellar variability.

When the two analyses are considered jointly, the models that show the most significant tension with the non-detection of binaries in the analysed sample are the reference model (R), together with NV and E5. In these cases, even if up to 70–80\% of the systems are consistent with the observed \ruwe\ within $3\sigma$, binary fractions higher than 0.6 are excluded by the Bayesian analysis in the metallicity range $-0.5<\mathrm{[Fe/H]}<-0.2$.
However, when chromaticity bias is included (V065), or when shorter or longer orbital periods are considered, the presence of a high binary fraction of at least $f_\mathrm{bin}\approx0.7$ cannot be excluded at high statistical significance, which may be consistent with $30\%$ contamination of the sample by metal-poor RR Lyrae.
This also shows that chromaticity bias should be carefully considered when interpreting the astrometry of variable stars in Gaia.

On the other hand, although a high-significance exclusion of the models cannot be claimed, there are several indications of possible tension between the models and the data.
First, the predicted \ruwe\ values in the binary models tend to be systematically higher than those observed. Second, the inferred binary fraction is below 0.1 at the $1\sigma$ level for all models, except for P03, where it remains below 0.5.
Finally, a few specific systems, most notably SW And and FW Lup (see Section~\ref{sec:ruwe}), are predicted to have very high \ruwe\ values, much higher than observed, in all models if they are assumed to be in binaries similar to those predicted by \cite{BI24}.
At the same time, the \ruwe\ analysis indicates that the overall best-fitting model is the one in which all RRLs are treated as single sources whose astrometry is affected by chromaticity-induced biases (SV065).

In conclusion, the astrometric capabilities of Gaia DR3 are not sufficient to interpret the lack of binary detections for metal-rich RRLs on thin-disc-like orbits as a rejection of the binary-formation model of \cite{BI24} at high statistical significance.
However, the results already indicate a tension that could plausibly be resolved either by considering a population largely dominated by single RRLs whose astrometry is significantly affected by chromaticity-induced biases, or by invoking a population that includes a non-negligible fraction of binaries with orbital periods substantially different from those predicted by the RRL binary-formation model, specifically, systems on longer orbits ($\sim 2000$–3000 days) or shorter ones ($\sim 300$–500 days).

We refrain from interpreting the possible tensions in Gaia DR3 as a clear constraint on the binary-formation channel for metal-rich RRLs. However, these tensions, if confirmed by a continued lack of detections in Gaia DR4 (Section~\ref{sec:dr4}), could point to the need for a deeper exploration of the uncertainties in stellar and binary evolution, for which metal-rich RRLs represent a particularly sensitive testing ground (see Section~\ref{sec:impact:rrl}).

\subsection{Comparison with other works}

In a companion study, \cite{Pranav26} perform a complementary population-level analysis by combining the binary-formation model of \citet{BI24} with Galactic population models \citep{Robin03,Vos20} and simulated Gaia observations generated with \texttt{gaiamock}. Using a setup comparable to our NV (no variability) model, they predict that $\approx 200$ out of $\approx 2600$ binary-formed RRLs within 3 kpc should have been detected in Gaia DR3, corresponding to a detectability of $\sim 0.07$. Although lower than our mean NV-model detectability ($\approx 0.17$), this difference is expected given the stricter quality cuts adopted in our sample.
In our case, accounting for chromaticity effects reduces the mean detectability by a factor of $\approx 1.5$. Applying the same analysis to the mock catalogue of \cite{Pranav26} yields $\approx 100$–130 detectable sources, consistent with this level of reduction.
Assuming orbital periods shorter by a factor of 0.3 leads to an even stronger reduction, yielding an expectation of $\approx 30$–40 detections.
Although the absence of detections in the specific subsample analysed here remains marginally compatible with their population-level predictions, the combination of the two works points toward the presence of a possible data–model tension (not limited to RRLs). This suggests that both a revision of the binary evolution models (see Section~\ref{sec:implications}), a reassessment of the population predictions from the adopted Galactic models (e.g. \citealt{Robin03}) and a more detailed analysis of the Gaia pipeline may be required.

The light-travel time effect, which produces long-term phase modulation in the light curves of variable stars in binary systems, has been widely used to search for RRL binaries. A systematic analysis by \citet{Hajdu21} identified 87 LTTE candidates with orbital periods in the range $\sim1000$–$4000$ days. Although none of these candidates overlap with our sample, preventing a direct comparison, their inferred period distribution is fully consistent with the orbital periods predicted by the binary-formation model of \citet{BI24} (see their appendix A).
A more targeted LTTE analysis of metal-rich, thin-disc RRLs was presented by \citet{Abdollahi2025}, who analysed 24 objects, 9 of which are in common with our sample.

For these stars, our models predict a mean Gaia DR3 detectability of $\langle \mathcal{O} \rangle \approx 0.5$ in the reference case, decreasing to $\approx 0.4$ when strong chromaticity effects are included (V065) and to $\lesssim 0.2$ when shorter orbital periods are assumed.
Only one object, ST~Pic, shows evidence of LTTE in that study. This source is not present in the analysed sample, but it is included in the alternative sample \texttt{ThinDisc} presented in Appendix \ref{app:alternative} and discussed in Section~\ref{sec:disc:cuts}. Notably, this source also has one of the highest predicted Gaia detectabilities in most of our models, given its close distance ($\approx 500$ pc). However, it does not exhibit any indication of anomalous astrometry ($\mathrm{RUWE} \approx 1$).
The coexistence of a possible LTTE signal and the absence of a Gaia DR3 astrometric detection may point to additional systematics not captured by our  \texttt{gaiamock} simulations, or to orbital periods shorter or longer than those predicted by the binary-formation model. However, any firm conclusion is limited by uncertainties in both our modelling and the LTTE analysis presented by \citet{Abdollahi2025}.

\rev{In particular, the authors do not report any specific estimate of the orbital period for this star, noting that a possible binary LTTE signal appears only as a secondary long-term variation superimposed on the more dominant modulation due to the Blazhko effect.}
Nevertheless, ST~Pic remains an intriguing object that warrants dedicated follow-up observations.

Proper-motion anomaly searches offer a further, complementary constraint. \citet{Kervella19a} analysed RRLs using the Hipparcos–Gaia DR2 baseline, identifying 190 candidates, 6 of which overlap with our sample. For these stars, our reference model predicts $\langle\mathcal{O}\rangle\approx0.68$, decreasing to $\approx0.5$ when chromaticity is included.
The proper-motion anomaly analysis of \citet{Kervella19a} has limited sensitivity to periods $\lesssim 1000$ days, so it is unlikely that the detected signal is consistent with shorter periods, as in model P03. Assuming longer periods and including chromaticity bias (V065P2), the detectability decreases to $\langle \mathcal{O} \rangle \approx 0.40$, marginally consistent (within 2--3$\sigma$) with the lack of Gaia detections (Equations \ref{eq:Ndet} and \ref{eq:Nvariance}).

\rev{\cite{Salinas26} detected 10 wide companions to RRLs via speckle interferometry at separations of $\sim$20--200 AU, much wider than the systems explored in this work. None of their detected binaries overlap with our analysed samples, but for the nine sources recoverable in the Gaia DR3 RRL catalogue, we ran dedicated \texttt{gaiamock} \ruwe estimates using the component properties reported by \cite{Salinas26}. The predicted \ruwe values for wide binaries match the observations better than the single-star case, independently confirming their detections and validating the sensitivity of our method.
\cite{Salinas26} also report an overall deficiency of wide binaries among RRLs relative to main-sequence stars, finding a $\sim$6\% fraction for the thin-disc population and $\sim$13\% for the most metal-rich RRLs, and suggest this casts doubt on the binary-formation channel of \cite{BI24}. However, at high metallicity, single-star evolution cannot produce RRLs within a Hubble time \citep{Zhang25}, so the expected number of metal-rich RRLs in wide binaries from single-star channels is effectively zero. Moreover, binary stars can interact only at separations of at most a few AU and are not expected to end up in the \citet{Salinas26} sample. The detections may instead trace non-interacting binary companions to classically-made metal-poor RR Lyrae, or higher-order multiplicity for binary-made RR Lyrae. Adopting the triple fraction from \cite{Moe19}, a moderate fraction ($\sim$13\%) of systems could host a binary-channel RRL alongside a wider tertiary, qualitatively comparable to the $\sim$6--13\% rate from \cite{Salinas26}. Their results are therefore not necessarily in tension with metal-rich RRLs originating from a binary-formation channel, even if the same possible tensions highlighted in this work apply as well to the possible inner binaries in the \cite{Salinas26} sample.}

Taken together, and within the uncertainties affecting both our analysis and previous studies, our results are qualitatively consistent with the existing literature. A coherent picture emerges in which models allowing for shorter or longer orbital periods (P03, P2 and V065P2) better reconcile the absence of Gaia DR3 astrometric detections, population-level expectations \citep{Pranav26}, the scarcity of LTTE detections among metal-rich RRLs \citep{Abdollahi2025}, and the proper-motion anomaly results of \citet{Kervella19a}.

\subsection{Predictions for DR4} \label{sec:dr4}

\begin{figure}
    \centering
    \includegraphics[width=\linewidth]{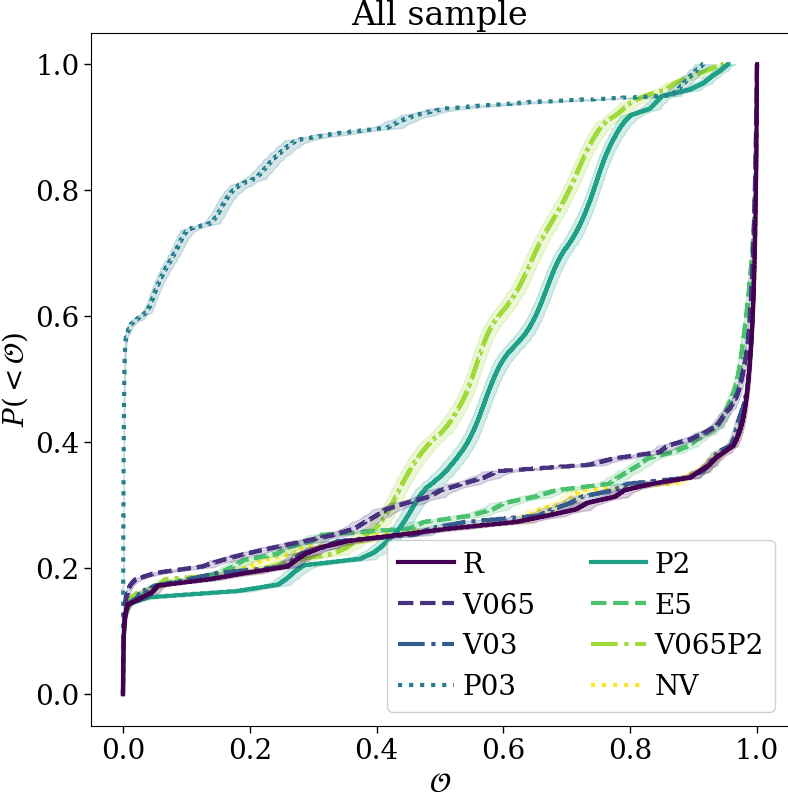}
    \caption{Same as the left panel of Figure \ref{fig:detdr3}, but considering the detectability in Gaia DR4.}
    \label{fig:detdr4}
\end{figure}

\begin{figure*}
    \centering
    \includegraphics[width=\linewidth]{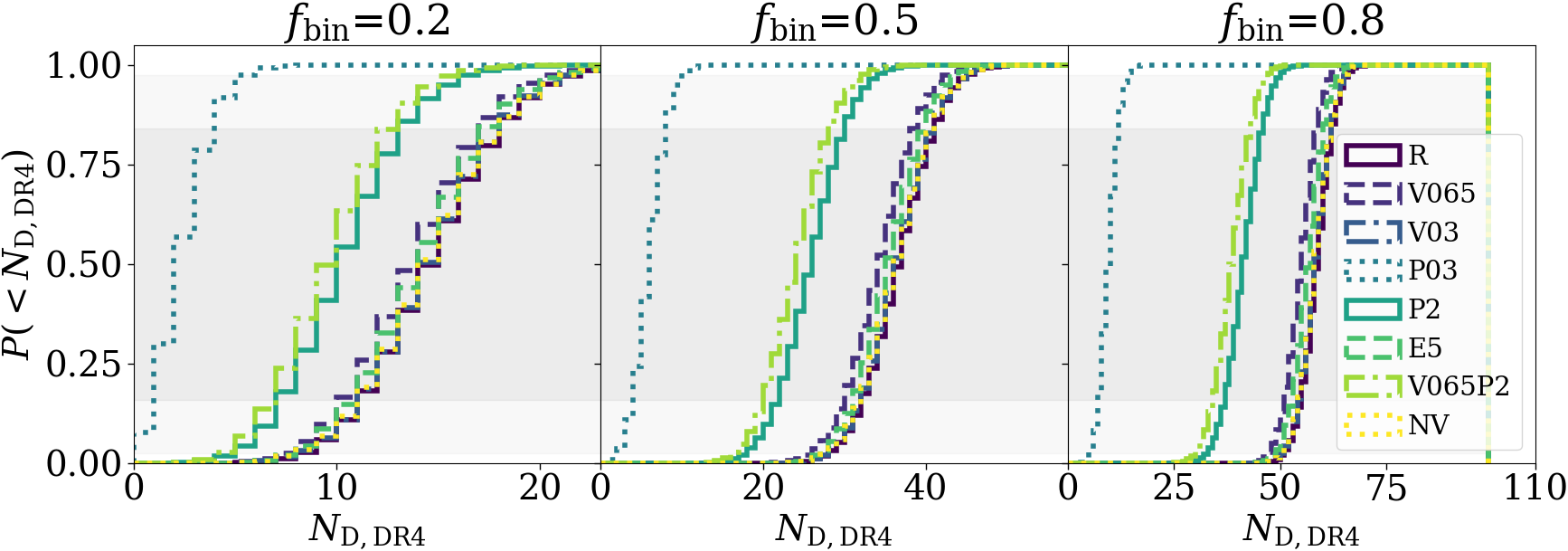}
    \caption{Cumulative distribution of the predicted number of astrometric binary detections in Gaia DR4 for the analysed RRL sample assuming different binary fractions ($f_\mathrm{bin}=0.1$, left panels; $f_\mathrm{bin}=0.3$, middle panels; $f_\mathrm{bin}=0.5$, right panels). Coloured curves correspond to the different models listed in Table~\ref{tab:models}. The gray shaded regions indicate the intervals centred on the median that enclose 84\% (dark gray) and 95\% (light gray) of the distribution.}
    \label{fig:Ndr4}
\end{figure*}

Assuming that Gaia DR4 will be produced using the same non-single-source pipeline as DR3, we estimate the binary detectability for the analysed sample of 100 RRLs (see Section \ref{sec:detectability}).

Figure~\ref{fig:detdr4} shows that, for most models, the detectability increases significantly compared to Gaia DR3. For the reference model (R), more than 60\% of the sample would be detectable with high confidence ($\mathcal{O}\approx1$). 
The chromaticity bias becomes less significant for detectability, even when assuming the model with the strongest impact (V065).
Models with longer orbital periods (P2 and V065P2) also become easier to detect, owing to the extended observational baseline provided by Gaia DR4. 
The short-period model (P03) remains the most pessimistic case, with about 80\% of the sample having $\mathcal{O}<0.2$, however a small subsample of six sources (FW Lup, SW And, HU Cas, CN Lyr, TW Her, TYC 9482-15-1) are predicted to be highly detectable in Gaia DR4 ($\mathcal{O}>0.85$).
For all models, essentially all detections ($\sim 100\%$) are obtained through full orbital solutions.

Figure~\ref{fig:Ndr4} shows the cumulative posterior distribution of the expected number of Gaia DR4 astrometric binary detections, $N_\mathrm{D,DR4}$ under different assumption on the binary fraction (see Appendix \ref{app:fitc}, and Equations \ref{eq:Ndet} and \ref{eq:Nvariance}).
 Figure~\ref{fig:Ndr4} shows that, owing to the substantially higher detectability in Gaia DR4, even a modest binary fraction ($f_\mathrm{bin}\sim0.2$) would result in tens of expected detections for almost all the model. 
The only scenario predicting significantly fewer detections is the short-period model (P03), for which the expected number of detections remains consistent with zero up to $f_\mathrm{bin}\sim0.8$.

The release of Gaia DR4 will provide a decisive test of the presence of binaries among RRLs with the properties predicted by binary-formation models. If no detections are obtained, and aside from possible changes in the Gaia processing or unknown systematics (Section~\ref{sec:systematics}), only models predicting significantly shorter orbital periods would remain compatible with a substantial population of hidden binaries. Such an outcome would have important implications for binary evolution theories (Section~\ref{sec:implications}).

These constraints are not limited to the metal-rich population. The models of \citet{BI24} predict, with strong confidence, that the old, metal-poor “classical” RRL population (e.g. in the stellar halo) should exhibit a binary fraction of order $f_\mathrm{bin}\sim0.05$, with orbital periods in the range $500$–$3000$ days. This assumes a parent halo binary fraction of $45\%$ (see \citealt{Abt83,Moe19}), with about $10\%$ of binaries RRLs having relatively short orbital periods  ($\lesssim 3000$ days).
These halo RRLs are expected to be non-interacting systems and should therefore not be affected by uncertainties or systematics related to binary evolution (see Section~\ref{sec:implications}). Their detectability instead depends primarily on systematics in the data processing pipeline.

For a complementary halo sample selected using the same criteria described in Section~\ref{sec:sample}, but adopting a metallicity cut at [Fe/H]$<-1.2$, we expect a total of approximately $1500$ stars. Assuming the mean Gaia DR4 detectability of the V065 ($\langle \mathcal{O} \rangle\approx 0.5$) and V065P2 ($\langle \mathcal{O} \rangle\approx 0.7$) models, which cover the expected period range, we qualitatively predict of order several tens ($30$--$50$) of detections in Gaia DR4 for a binary fraction $f_\mathrm{bin}\sim0.05$ (Equation~\ref{eq:Ndet}).

Unless driven by specific choices in the Gaia non-single-star pipeline, a global absence of binary RRL detections in DR4 would pose a challenge extending well beyond the formation of metal-rich RRLs. Such an outcome would point either to more significant variability-induced astrometric biases than those captured by our models, to additional instrumental or pipeline effects not included in our mock procedure, or to tensions with current estimates for the initial properties and fraction of Sun-like binary systems (see, e.g., \citealt{Moe19}). It could also reflect uncertainties in the metallicity-dependent maximum radial expansion during the RGB phase, which directly affects the likelihood of triggering binary interactions \citep{Piau11}, although this would imply an extreme production of other stripped horizontal branch stars, including sdB binaries, which are strongly constrained at such metallicities \citep{Molina26}.

\subsection{Implications} \label{sec:impact:rrl} \label{sec:implications}

The large uncertainties and systematic effects (Section~\ref{sec:systematics}) affecting both the data and the modelling do not allow us to statistically exclude, at high significance, the possibility that the binary-formation model presented by \citet{BI24} is consistent with the absence of Gaia detections in the analysed sample.
Nevertheless, the nominal outcome of our investigation points to a tension between the observations and the model predictions.

Based on these results, we discuss the possible implications for the formation channels of metal-rich RRLs and, more generally, for binary evolution models, for which this population represents a directly testable prediction.

\subsubsection{Variation of the binary-formation channel}

The orbital predictions of the \citet{BI24} model rely on specific assumptions about mass and angular-momentum loss during mass transfer. In particular, the transfer is assumed to be highly non-conservative, with the material transferred to the companion being expelled from its vicinity and removing the companion’s specific orbital angular momentum \citep{Vos20}. 
This configuration in their study is motivated by the orbital properties of observed long-period hot subdwarf stars (sdBs), which depend sensitively on the degree of mass transfer conservativeness. In contrast, as discussed in \citet{Vos20}, the properties of these binaries are much less sensitive to the assumed angular momentum loss prescription. Therefore, the latter remains an important uncertainty in the binary models. 

Studies of sdB from more massive B-type progenitors than modelled in \citet{Vos20} suggest that mass transfer, at least for a different population of more massive stars, may be more conservative for those masses \citep{Lechien25}. Mass may also be lost as a wind from the donor \citep{Olejak25}, or through the outer Lagrangian points, or via a circumbinary disc \citep{Marchant21}. If such processes can change the properties of metal-rich RRLs while keeping the predictions about hot subdwarf binaries unchanged, it could provide a viable explanation for the tension seen here. 

In Appendix~\ref{app:angmom}, we discuss that the \citet{BI24} prescription may lead to the longest final periods, at least within a certain class of angular momentum loss models. 
Furthermore, even a modest fraction of mass lost through the equatorial plane (of order $\sim20\%$) can potentially shrink the orbit significantly, yielding final periods of $\sim200$–$500$ days, consistent with model P03, or the orbits may also expand significantly if a gravitationally bound circumbinary disc is formed and is heated by the central star \citep{Heath20}. The circumbinary disc models are particularly interesting because sdB binaries are likely to remove the disc faster than RRL progenitors, making it more important for RRLs. The short-period, $\sim200$–$500$ days, region of parameter space is also observationally critical, since such systems are typically too wide to produce eclipses or clear Gaia astrometric solutions, yet too compact for efficient LTTE detection. Their radial-velocity semi-amplitudes ($\sim20$–$40 \ \mathrm{km \ s^{-1}}$) are comparable to RRL pulsational velocities, complicating spectroscopic identification. Circumbinary discs may also help explain peculiar surface abundances and eccentric post-mass-transfer systems \citep{Vos15,Moltzer25,Martin25,DOrazi24}, and the mean-magnitude variations observed for some binary candidate RRLs in the MW bulge regions \citep{Hajdu26}.

All models considered here assume initially circular binaries. While sdB binaries are well known to have eccentricities up to $0.5$ \citep{Molina26}, circular orbits were used partly because of the lack of tools to model eccentric mass transfer and partly because of the yet unknown nature of the eccentricity. While tidal circularisation is commonly assumed, incomplete synchronisation, residual eccentricity, the interplay of eccentricity and mass transfer, or tidal toques from the circumbinary disc may all contribute to eccentricity growth \citep{Vos15,Hendriks23,Rocha25,Parkosidis25}. While moderate eccentricities up to $e\approx0.5$ do not significantly alter Gaia detectability in our analysis, there seems to be an empirical preference for Gaia to detect the higher-eccentricity part of the population, at least for sdBs \citep{Pranav26, Molina26}. 
Eccentric binaries may arise naturally in dynamically active environments, where RRL progenitors could form eccentric binaries with compact objects or participate in higher-order multiple systems. In particular, triple-induced eccentricity excitation \citep{Rastello25} could trigger mass transfer at large separations. The discovery of a metal-rich RRL in Trumpler~5 \citep{Mateu25} and the statistical association of RRLs with young and intermediate-age clusters in the Magellanic Clouds \citep{Cuevas24} further motivate the exploration of the influence of cluster dynamics.

A possible reason for the lack of Gaia DR3 detections is the modelled luminosity of the RRL companion. Specifically, \cite{Molina26} found that in long-period sdB binaries, the companion is observed to be almost  1 magnitude brigheter in the  Gaia $G$-band  than the \cite{Vos20} model predicts. If the RRLs companion main sequence stars are similarly brighter, it will reduce the motion of the photocenter and reduce the binary detectability in Gaia. Furthermore, both \citet{Vos20} and \citet{BI24} focused only on degenerately-ingiting progenitors with primary masses below approximately $2.1\,$M$_\odot$. These studies also imply (see also \citealt{Karczmarek17}) that progenitors with more massive primaries will produce shorter-period binaries, with periods between a few tens and a few hundreds of days. The full population of metal-rich RRLs will therefore contain both short- and long-period systems, which may help explain Gaia observations.

The adopted models also assume weak RGB winds, preventing single-star formation of RRLs with $\mathrm{[Fe/H]} > -1.0$. Stronger winds could reduce the mass that must be removed through binary interactions, although this is likely subdominant during active mass transfer  \rev{and a high wind efficiency at this metallicity regime is not consistent with asteroseismic observations \citep[][and references therein]{Yaguang25}}. 
Moreover, differences in stellar modelling, including opacity, mixing-length calibration, boundary conditions, and helium and $\alpha$-element abundances, affect the maximum RGB radius and the core–envelope structure, thereby potentially impacting both the onset and outcome of mass transfer (see e.g. \citealt{Aguirre20}).
An additional mechanism that may significantly affect the final orbital period is tidally enhanced stellar winds, whereby the presence of a companion boosts the red-giant mass-loss rate prior to Roche-lobe overflow \citep{Tout88}. This process can remove a substantial fraction of the envelope at relatively wide separations, potentially producing longer-period systems without requiring classical mass transfer, similar to models P2 and V065P2. Although invoked to explain horizontal-branch morphologies \citep{Lei13,Vos15}, it has not yet been explored specifically for partially stripped, metal-rich RRL progenitors, and there exist strong constraints on its viability \citep{Jorissen98, Vos15, Escorza20}. Additionally, it still remains highly challenging to produce a modified metal-rich RRL population through these modifications, while keeping the sdB predictions in agreement with observations.

Binary-made RRLs and stripped core-helium-burning sdB stars differ by only $\approx0.04$–$0.07\,\mathrm{M}_\odot$ of residual envelope mass \citep{BI24}, so variations in mass- and angular-momentum loss prescriptions, at face value, are expected to impact both populations similarly. Observationally, sdB binaries are found both at $\sim400$–$600$ days and at longer periods, and they exhibit a clear period–metallicity correlation \citep[][and references therein]{Molina26}. In contrast, the subclass of alternative prescriptions explored in Appendix~\ref{app:angmom}, which are capable of producing shorter-period binary-made RRLs, tend to erase such a correlation. Moreover, sdB binaries have already been detected astrometrically in Gaia, albeit in smaller numbers than predicted by the model \citep{Pranav26}. In this context, the absence of detected RRL binaries, especially classically made, metal-poor RRL, is particularly noteworthy. While variability-induced astrometric biases likely reduce RRL detectability, this effect alone appears only marginally sufficient to explain the lack of detections. However, realistic pulsation models, showing LTTE modulations, Blazhko effect, or mean-magnitude variations may all undermine Gaia's ability to characterise such systems.

\subsubsection{Metal-rich RR Lyrae stars as single sources} \label{sec:rrlsingle}

If a fraction of metal-rich RRLs in our sample are truly currently single, this still does not exclude a binary origin. In this case, past interactions may have proceeded through unstable mass transfer rather than the stable channel modelled by \citet{BI24}, leading instead to common-envelope evolution or mergers. Indeed, the models also show that a subset of interactions leads to unstable mass transfer \citep[e.g.,][]{Karczmarek17,Vos20,Iorio23,BI24}. The outcomes include highly stripped short-period post-common-envelope binaries and may include common-envelope mergers, also called failed common-envelope evolution.
Interactions involving a low-mass main-sequence star and a white dwarf, including common-envelope mergers, may also produce horizontal-branch stars, including RRLs, as was proposed in the scenario for undermassive red-clump stars \citep{Rui24,Matteuzzi23}, although there still needs to be a population study to determine its importance. The same scenario may also operate in stellar collisions, particularly in dense environments such as young massive clusters and globular clusters.

Purely single-star evolutionary channels appear disfavoured, as removing the large envelope mass required to form metal-rich RRLs ($\gtrsim 0.4,M_\odot$) through RGB winds alone is unlikely \citep{Yaguang25}. Additional mechanisms such as helium enrichment (e.g. \citealt{Marconi18}), rotation-enhanced winds, or episodic mass-loss events would be required. However, helium-enriched RRL models generally predict higher luminosities than observed, and RRLs in helium-enhanced globular clusters (e.g. NGC~6441, \citealt{Caloi07}) show longer periods and distinct pulsation properties compared to field RRLs at similar metallicity \citep{Reyes24}. Rapid rotation could, in principle, enhance winds and induce internal mixing \citep{Sweigart97}, but high rotation rates in evolved low-mass stars are themselves usually linked to binarity or mergers (see e.g. \citealt{Carlberg11}).

Overall, metal-rich RRLs represent a powerful probe of stellar and binary evolution, as their existence requires substantial mass loss at high metallicity and is sensitive to the details of binary interactions. In addition to the stable mass transfer channel, several other channels may be operating, including common envelope mergers, triple stars, and, in dynamical environments, stellar collisions, all of which are viable and require further exploration.

\subsection{Systematic Effects and Caveats} \label{sec:systematics}

\subsubsection{RRL samples and selection cuts} \label{sec:disc:cuts}

Our analysis is based on the non-detection of RRL binaries in Gaia DR3. As a result, the resulting constraints depend sensitively on the adopted sample size, reflecting the strong dependence of the likelihood on the number of sources (Equation~\ref{eq:lkl}).
Smaller samples naturally allow higher binary fractions to remain consistent with non-detections, whereas larger samples increasingly predict detections even when the average detectability of individual systems is low. For this reason, we focus on a sample that, according to the binary-formation model of \citet{BI24} and the results of \citet{Zhang25}, is expected to contain a high fraction of metal-rich RRLs with intermediate ages, for which binary formation appears to be the most likely evolutionary channel.

The astrometric-quality cuts derived from the \cite{Zhang25} and \citet{IB21} 
could in principle exclude binaries with degraded astrometric solutions because of binarity.
However, the purpose of this work is not to measure the global RRL binary fraction, but to assess whether a specific class of binaries predicted by the models could remain undetected in Gaia. 
Nevertheless, uncertainties in photometric metallicities and statistical kinematic classifications imply that some contamination by RRLs outside the targeted population cannot be excluded, which would affect the inferred constraints.

To test the impact of these systematics, we repeat the analysis using four alternative samples described in Appendix~\ref{app:alternative}, in which we explore alternative astrometric  and metallicity cuts and include a sample with robust high-resolution spectroscopic metallicities.
Overall, our analysis confirms that the main sample represents the best compromise between completeness and purity when selecting RRLs with kinematics consistent with intermediate-age thin-disc populations. At the same time, when accounting for uncertainties and systematics, the results obtained with the three alternative samples are broadly consistent with those from the main sample.
In particular, the analysis of the alternative sample \texttt{Plx04}, obtained by imposing a stricter parallax cut on the original sample, indicates that for the closest stars ($\lessapprox2.5$ kpc) a high binary fraction cannot be excluded with high statistical significance in any of the tested models. However, the selected subsample does not reduce the systematic overestimate of the \ruwe\ predicted by the binary models.

\subsubsection{Distance} \label{sec:disc:distance}

The source distance is a key factor for Gaia’s ability to detect binaries. In this work, distances are estimated by inverting the Gaia DR3 parallaxes, applying a global zero-point correction of $-0.033$ mas as measured for field RRLs by \citet{Garofalo22}, consistent with other Gaia DR3 estimates \citep{Groenewegen21}. We find excellent agreement (residuals centred on zero with  scatter below 100 pc) with the distances estimated with more sophisticated Bayesian distance estimates \citep{Bailer-Jones21,Weiler25}.

As an additional check, we compare our distances with those inferred from the $G$-band absolute magnitude–metallicity relation of \citet{Garofalo22}, following the procedure of \citet{IB21}. The two estimates agree well, with residuals centred on zero and a dispersion of $\sim300$ pc, consistent with the expected uncertainties after removing a number of outliers likely associated with biased metallicities or contaminants ($\lesssim 15 \%$, see e.g. \citealt{Ranaivomanana2025} and Appendix \ref{app:V552Car} for similar cases).
This estimated level of contamination is consistent with the $20$–30\% discussed in Section~\ref{sec:sample} and taken into account in the comparison between the data and the model expectations.

We also test the impact of omitting the parallax zero-point correction, which systematically increases distances by up to $\sim500$ pc at $\varpi=0.25$ mas. Repeating the analysis under this extreme assumption does not lead to significant changes in our results.

Finally, unresolved binarity could in principle bias parallaxes \citep{ElBadry25}. However, our mock catalogues indicate that parallaxes remain well recovered even in the presence of binary motion, with signal-to-noise ratios typically $\gtrsim 2$.
Overall, this suggests that for nearby objects, where the parallax signal is intrinsically strong, binary-induced astrometric systematics should primarily manifest as elevated \ruwe\ rather than as severely biased parallaxes with correspondingly inflated formal uncertainties.

These tests indicate that our distance estimates are sufficiently robust and that distance-related systematics do not significantly affect our conclusions.

\subsubsection{Metallicity estimate} \label{sec:metallicity}

Our results depend on the metallicity distribution of the analysed sample. We adopt photometric metallicities, which carry significant uncertainties and may be affected by systematics. Our estimates are consistent with recent state-of-the-art determinations by \cite{Muraveva25}, but tend to be systematically lower at the high-metallicity end compared to some other studies \citep[e.g.][]{Mullen21}. While the statistical uncertainties are fully propagated in our hierarchical Bayesian framework by treating the true $\mathrm{[Fe/H]}$ as a latent variable, a significant systematic offset could still affect the inferred constraints as a function of metallicity.

In the extreme case of underestimated metallicities, the metal-rich tail would be artificially under-populated, and a less biased sample would contain more high-$\mathrm{[Fe/H]}$ sources. This would tighten the constraints on $f_\mathrm{bin}$ at high metallicity by reducing the impact of small-number statistics, potentially increasing the tension with models predicting an enhanced binary fraction. We consider this scenario unlikely: the adopted threshold $\mathrm{[Fe/H]}=-0.5$ is motivated by \citet{Zhang25}, who use the same photometric metallicity calibration with no obvious mechanism to preferentially populate the metal-rich RRL regime.

As a further check, we repeat the analysis using the smaller but more conservative high-resolution spectroscopic RRL sample from \citet{DOrazi24}. Owing to its limited size (71 objects after cuts, with only six at $\mathrm{[Fe/H]}>-0.5$), the resulting constraints are much weaker and do not exclude high binary fractions for any of the tested models (Appendix~\ref{app:alternative}).

\subsubsection{Gaia mock observation} \label{sec:gaiasim}

The tendency of the mocks to produce slightly higher eccentricities \citep{ElBadry24} does not affect our conclusions, since we verify that eccentricities up to $e=0.5$ have a negligible impact on our results, and the adopted binary-formation model assumes circular orbits.
\citet{ElBadry25} also noted that \texttt{gaiamock} tends to slightly overpredict the \ruwe\ of binaries at low values (see their Figure 1), but at a level that is completely negligible for our analysis (median difference $\lesssim 0.02$).

On the other hand, metal-rich RRLs, if formed through stable mass transfer, are strongly expected to be eccentric, given that hot subdwarfs are known to be eccentric. Moreover, due to the positive overall period-eccentricity correlation for post-mass-transfer binaries, the RRL eccentricities can be of the order $e\approx0.2 - 0.5$ if they form at orbits of $1000$ days. As Gaia tends to detect the less eccentric sub-population tails of binaries, eccentricity may have an important effect on their detectability, especially for binaries with the longest periods (and hence the highest eccentricities). Binary classical metal-poor RRLs are also likely to be comparably eccentric based on the observed properties of their main-sequence progenitors.

A key source of uncertainty concerns variability-induced astrometric biases, in particular chromaticity effects. In our framework, chromaticity is treated with a simplified prescription, whereas in Gaia it is handled through a much more complex calibration pipeline. Although our model reproduces the observed correlation between \ruwe\ and RRL pulsation amplitude, the true strength of the effect remains uncertain. If overestimated, chromaticity would play a minor role and the reference model would be more appropriate, increasing the tension with observations; if underestimated, detectability could be suppressed even more strongly than in our conservative models.
In addition, in our model, the chromatic effect is implemented as a pure systematic shift along the scan direction, and we do not consider any variability-induced inflation of the astrometric uncertainties. An underestimate of the astrometric errors would therefore likely lead to an inflation of the \ruwe. Although we find it unlikely that such a bias, if present, could, by itself, account for the \ruwe\ systematic offset between binary models and the data, it could nevertheless help alleviate the discrepancy.
In this context, it would be interesting to extend this analysis to other stellar populations which are known to be variable but whose formation does not, in principle, require them to be in a close binary system at present (for example, Delta Scuti stars, \citealt{Catelan15}).

Future Gaia DR4 data, including epoch astrometry for all sources and access to raw image data, will enable a more detailed characterisation of variability-induced astrometric biases and provide a critical test of the assumptions adopted in this work.

\section{Summary and conclusions} \label{sec:summary} 

We investigated whether the absence of astrometric binaries hosting RR Lyrae stars (RRLs) in Gaia DR3 is compatible with binary evolution models proposed to explain the origin of metal-rich and intermediate-age ($<9$ Gyr) RRLs in the Milky Way disc. These models predict that most metal-rich and intermediate-young RRLs form through stable binary mass transfer and reside in binaries with orbital periods of $\sim$900--2000 days, a region of parameter space that Gaia DR3 is in principle sensitive to. We adopted a forward-modelling approach combining a carefully selected sample of 100 metal-rich RRLs with kinematics consistent with an intermediate-young disc population \citep{Zhang25}, detailed binary evolution predictions, and realistic simulations of Gaia astrometric observations. Using an updated version of the \texttt{gaiamock} framework \citep{ElBadry24mock}, we accounted for variability-induced astrometric effects including heteroskedastic errors, variability-induced motion, and chromaticity biases, exploring a range of assumptions for orbital periods, eccentricities, and variability systematics.

Our main results can be summarised as follows:

\begin{itemize}

\item Considering the model and data uncertainties and systematics, we cannot rule out at high statistical significance ($99\%$ confidence level) that the lack of RRL binaries in Gaia DR3 for the analysed sample is consistent with a high fraction of hidden binaries ($f_\mathrm{bin} \gtrsim 0.7$). Although no firm conclusion can yet be drawn, the comparison between data and models already suggests a possible tension. Most notably, the \ruwe\ values predicted by the models tend to be systematically higher than those observed, and, at relatively modest statistical significance ($\lesssim$$95\%$), most models already disfavor binary fractions larger than 0.5.

\item The tension is significantly reduced only for models predicting shorter orbital periods ($\sim$300--500 days), or longer orbital periods ($\sim$2000--3000 days) when variability-induced astrometric biases are also taken into account.

\item We show that variability-induced astrometric biases must be taken into account when interpreting high \ruwe\ values in variable sources. These effects can both inflate the observed \ruwe\ distribution and reduce the probability of accepting astrometric binary solutions. Interestingly, a scenario in which all RRLs in the sample are single, with their \ruwe\ distribution entirely dominated by variability-induced biases, provides the best overall match to the data.

\item In Gaia DR4, the expected detectability will improve significantly, with tens of detections anticipated for the analysed sample even assuming a low intrinsic binary fraction ($f_\mathrm{bin} \approx 0.2$). Gaia DR4 will thus have the capability to provide a strong test for the binary evolution models. Assuming a perfectly accurate detection pipeline, the only models that will remain consistent with zero detections will be those with significantly shorter orbital periods in the range 300--500 days.

\item A number of RRLs are predicted to have very high binary detectability ($>0.85$) in Gaia DR4, even under the most pessimistic assumptions. These are: FW~Lup, SW~And, HU~Cas, CN~Lyr, TW~Her, and TYC~9482-15-1. The absence of detections among these objects will place the strongest constraints on their formation models and on the Gaia detection pipeline for binary pulsators, making them key targets for follow-up observations.

\end{itemize}
Our results reveal a possible tension between existing binary evolution predictions and the absence of astrometric detections in Gaia DR3. A continued absence of detections in Gaia DR4 would not necessarily exclude the binary-formation channel, but would motivate a broader exploration of detection uncertainties and binary evolution assumptions, with implications for stellar-population studies. Metal-rich RRLs therefore emerge as sensitive probes of key phases of binary evolution. As part of this work, we developed and publicly released an extension of the \texttt{gaiamock} framework incorporating variability-induced astrometric biases, providing a community tool for forward modelling Gaia observations of variable stars.

\section*{Data availability}

All the data and the scripts to reproduce the results and analysis of the paper are avaiable at \url{https://doi.org/10.5281/zenodo.18499248}.
The custom version of the \texttt{gaiamock} tool developed for this work is available at \url{https://github.com/giulianoiorio/gaiamock} as a forked version of the original repository (\url{https://github.com/kareemelbadry/gaiamock}).

\begin{acknowledgements}
\rev{The authors thank the anonymous referee for their suggestions that helped us to improve the manuscript.}.
GI, AB, EP, VD, CM acknowledge the monthly Synergy meeting initiative and the member of the Synergy group for useful discussion and feedbacks. 
GI thanks the members of the Dynamics Group, Gaia Group (in particular Francesca Figueras), and the ERC-starting grant-funded group CET-3PO (in particular Nadejda  Blagorodnova) at the University of Barcelona for useful discussion and feedbacks.
GI thanks the friends of \emph{Fantaporcheria}, whose company and humour made even the most intense days feel lighter.
GI was supported by a fellowship grant from the la Caixa Foundation (ID 100010434). The fellowship code is LCF/BQ/PI24/12040020.
GI, SR and MG acknowledge the grants PID2024-155720NB-I00, CEX2019-000918-M, CEX2024-001451-M funded by MCIN/AEI/10.13039/501100011033 (State Agency for Research of the Spanish Ministry of Science and Innovation). AB acknowledges support from the Australian Research Council (ARC) Centre of Excellence for Gravitational Wave Discovery (OzGrav), through project number CE230100016. HZ thanks the Science and Technology Facilities Council (STFC) for a PhD studentship (grant number 2888170). PN and KE acknowledge support from NSF grant AST-2307232. 
Co-funded by the European Union (ERC-2022-AdG, "StarDance: the non-canonical evolution of stars in clusters", Grant Agreement 101093572, PI: E. Pancino). Views and opinions expressed are however those of the author(s) only and do not necessarily reflect those of the European Union or the European Research Council. Neither the European Union nor the granting authority can be held responsible for them.
SR acknowledges financial support from the Beatriu de Pinós postdoctoral fellowship program under the Ministry of Research and Universities of the Government of Catalonia (Grant Reference No. 2021 BP 00213). VD acknowledges partial support from the INAF Minigrant 2024 {\it MUGS}. 
This research was supported in part by grant NSF PHY-2309135 to the Kavli Institute for Theoretical Physics (KITP). Part of this research was carried on during the KITP long program \emph{Stellar-Mass Black Holes at the Nexus of Optical, X-ray, and Gravitational Wave Surveys}
This work has made use of data from the European Space Agency (ESA) mission
{\it Gaia} (\url{https://www.cosmos.esa.int/gaia}), processed by the {\it Gaia}
Data Processing and Analysis Consortium (DPAC,
\url{https://www.cosmos.esa.int/web/gaia/dpac/consortium}). Funding for the DPAC
has been provided by national institutions, in particular the institutions
participating in the {\it Gaia} Multilateral Agreement.
This research has made use of the tool provided by Gaia DPAC to reproduce the Gaia (E)DR3 photometric uncertainties described in the GAIA-C5-TN-UB-JMC-031 technical note using data in Riello et al. (2021).
\rev{We thankfully
acknowledge the computing resources provided by the Red Espa\~{n}ola de Supercomputación (RES) under the projects AECT-2024-3-0023, AECT-2025-1-0029, AECT-2025-2-0037, and
AECT-2025-3-0014. This includes access to the PIRINEUSIII supercomputer at Consorci de Serveis Universitaris de Catalunya (CSUC), and the MARENOSTRUM5
at Barcelona Supercomputing Center (BSC). GI
acknowledges the use of the NYX supercomputing cluster at ICCUB.}
\end{acknowledgements}

\bibliographystyle{aa_edited} 
\bibliography{main}

\begin{appendix}

\section{RR Lyrae stars in the Gaia DR3 binary catalogue} \label{app:gaiarrlbin}

\begin{figure}
    \centering
    \includegraphics[width=\linewidth]{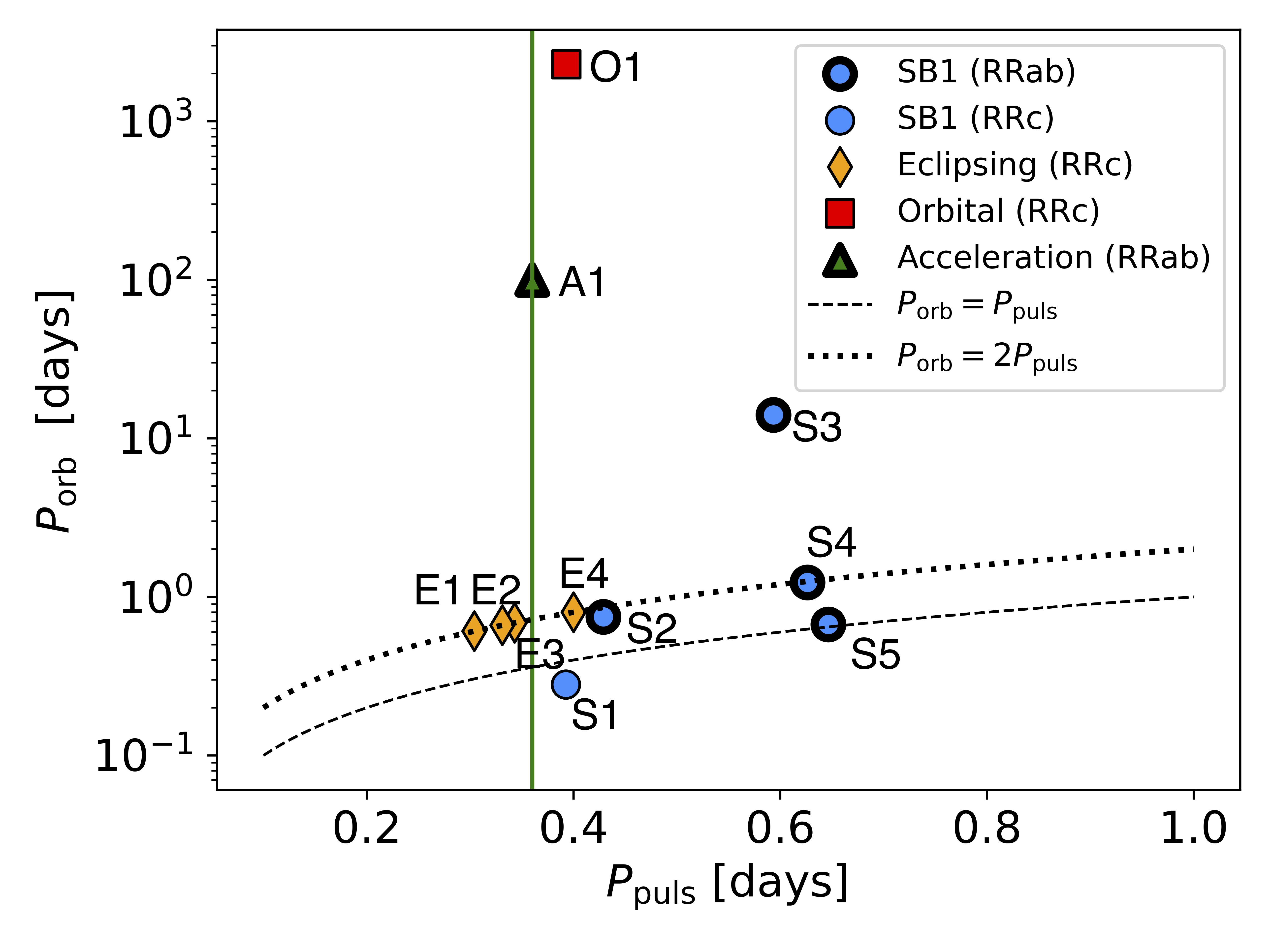}
    \caption{RRL pulsation period vs orbital period for Gaia DR3 sources in the non-single source DR3 catalogues. Blue circles: single-line spectroscopic binaries (S); yellow diamonds: eclipsing binaries (E); red square: full astrometric orbital solution (O); green triangle: non-linear proper motion solution (A; no orbital period available). Thick (thin) contours indicate RRab (RRc) stars. Dashed and dotted lines mark $P_{\rm orb}=P_{\rm puls}$ and $P_{\rm orb}=2P_{\rm puls}$, respectively.}
    \label{fig:rrlbindr3}
\end{figure}

\begin{figure*}
    \centering
    \includegraphics[width=\linewidth]{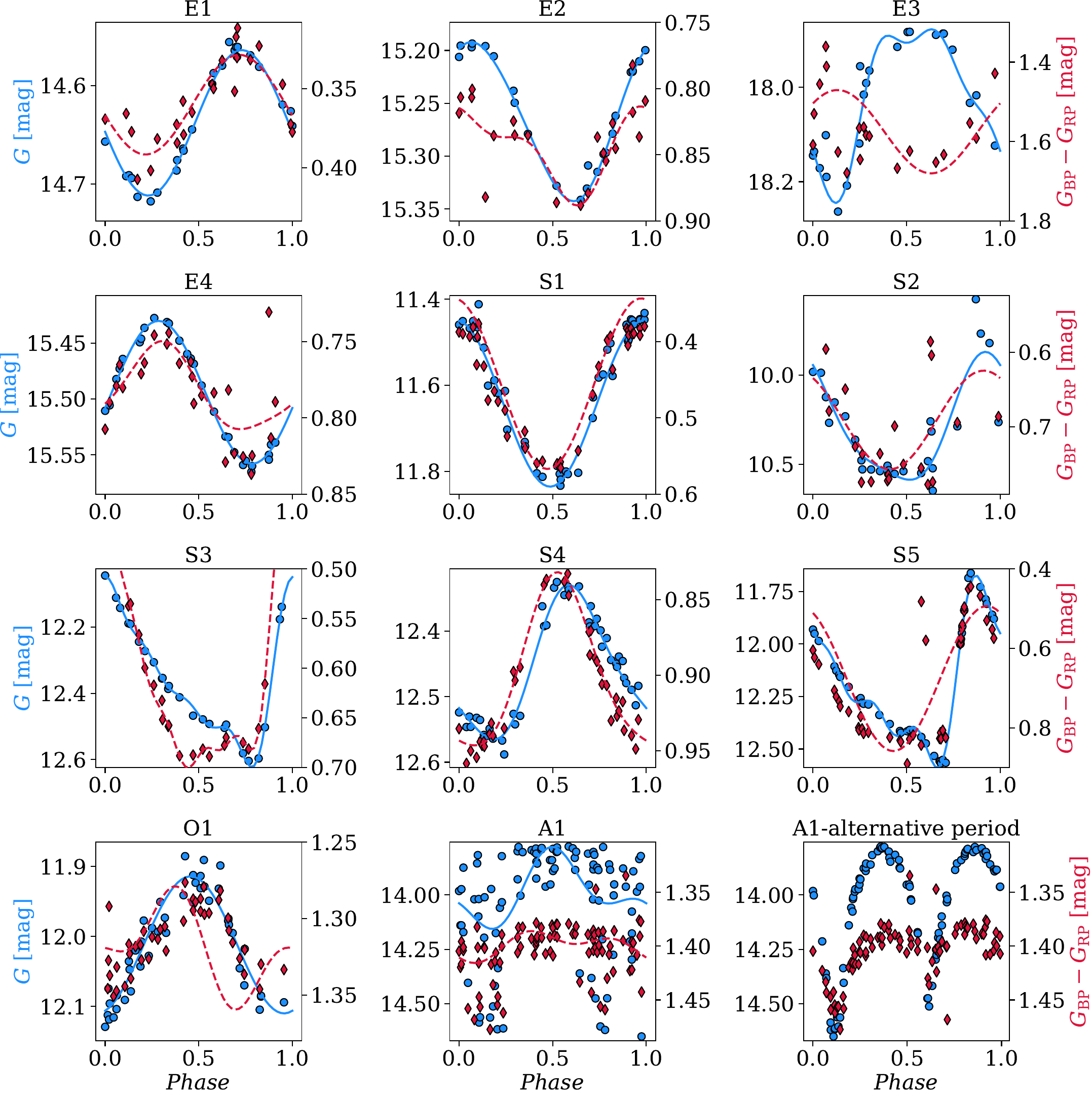}
    \caption{Phase-folded Gaia DR3 epoch photometry in the $G$-band (blue circles; left y-axis) and colour ($G_\mathrm{BP}-G_\mathrm{RP}$; red diamonds; right y-axis) for the ten Gaia DR3 RRL candidates found in the non-single-star Gaia DR3 catalogues. The phase folding is performed using the Gaia DR3 RRL pulsation period. Curves show the reconstructed harmonic light-curve model from the Gaia DR3 catalogue (solid blue for $G$; dashed red for the colour). Each panel corresponds to a single object, labelled according to the nomenclature adopted in Figure~\ref{fig:rrlbindr3}. The bottom-right panel (A1-alternative period) shows the phase-folded light curve of A1 using the alternative period of 0.296~days (see Appendix \ref{sec:aone}).}
    \label{fig:rrlbindr3lc}
\end{figure*}

In Gaia DR3, there are eleven sources that appear both in the RRL catalogue \citep{Clementini23} and in one of the four non-single-star catalogues \citep{GaiaDR3binary}. Figure~\ref{fig:rrlbindr3} shows their location in the orbital-period versus variability-period plane, while Figure~\ref{fig:rrlbindr3lc} presents the phase-folded Gaia DR3 epoch photometry together with the light-curve models reconstructed from the Gaia DR3 harmonic decomposition.
Ten of these objects are listed in the \texttt{nss\_two\_body\_orbit} catalogue: four are classified as eclipsing binaries (E), four as single-line spectroscopic binaries (S), and one has a full orbital astrometric solution (O). The remaining object is instead found in the \texttt{nss\_acceleration\_astro} catalogue, which contains sources showing evidence of non-linear proper motions (A). No objects are found in the \texttt{nss\_vim\_fl} catalogue, which contains VIM solutions (see Appendix~\ref{app:vim}).
\cite{Pranav26} reports that all these sources correspond to spurious detections (either misclassified RRLs or misclassified non-single-star solutions). In this Appendix, we confirm these results and complement them with a discussion of the individual systems.

\subsection{Eclipsing binaries, \Eone, \Etwo, \Ethree, \Efour} \label{app:eclipse}

All the RRLs classified as eclipsing binaries are RRc stars and have orbital periods close to twice their pulsation periods. This is precisely the relation expected in cases of misclassification between RRLs and eclipsing binaries. In particular, the nearly sinusoidal light curves of RRc stars can mimic those of contact eclipsing binaries (W Ursae Majoris type), which are in fact the dominant source of contamination in RRL samples \citep{Rimoldini23}.
Figure~\ref{fig:rrlbindr3lc} shows that for \Eone (Gaia DR3 3669003373413914624), \Etwo (Gaia DR3 4073896567545411328), and \Efour (Gaia DR3 4207618935502749184) the colour variations follow the expected RRL behaviour, with the sources becoming bluer near maximum light and redder at minimum light. For \Eone and \Etwo, folding the data on the orbital period results in a noisier trend, supporting the interpretation that these objects are genuine RRLs with a spurious eclipsing-binary classification. In contrast, \Ethree (Gaia DR3 4285576371518811776) shows irregular colour variations and a light-curve morphology not fully consistent with an RRc star; when folded on the orbital period, the modulation appears more regular and consistent with an eclipsing binary. We therefore interpret \Ethree as a likely eclipsing system misclassified as an RRc star.

\subsection{Spectroscopic binaries, \Sone, \Stwo, \Sthree, \Sfour, \Sfive}

Figure~\ref{fig:rrlbindr3lc} shows that all the RRLs classified as single-line spectroscopic binaries display light curves broadly consistent with genuine RRL variability, with the only partial exception of \Stwo (Gaia DR3 6771307454464848768), which appears noisier, particularly in the colour trend. The reported radial-velocity semi-amplitudes correlate with the photometric amplitudes and are consistent with the empirical relations for RRab and RRc stars reported by \citet{Prudil24}. For \Sone (Gaia DR3 1193137104465445888) and \Sfive (Gaia DR3 5764934181067920256), the reported orbital periods are comparable to the pulsation periods, while for \Stwo and \Sfour (Gaia DR3 6032843799994897408) they are close to twice the pulsation period. Overall, these properties suggest that these objects are likely genuine RRLs, for which the pulsation-driven expansion/contraction cycle has been erroneously interpreted as orbital radial-velocity variability.

\subsubsection{\Sthree (CV~Scl)}

The case of \Sthree (also known as CV~Scl) is intriguing. It is the only object in the SB1 sample for which the reported orbital period ($14.04$ days) shows no obvious connection to the pulsation period ($0.59$ days). The proposed binary orbit has a low eccentricity ($e = 0.17 \pm 0.12$). The reported radial-velocity semi-amplitude is $29.2 \pm 2.7 \ \mathrm{km \ s^{-1}}$. This value is smaller than expected from pulsation alone, given the \Sthree\ light-curve amplitude, which would suggest $\approx 55 \ \mathrm{km \ s^{-1}}$.

A back-of-the-envelope estimate based on the parallax gives an absolute magnitude of $M_\mathrm{G} \simeq 0.32$. Accounting for the parallax offset (Section~\ref{sec:mock}), this becomes $M_\mathrm{G} \simeq 0.48$. These values correspond to $\log(L/L_\odot) \simeq 1.72$–$1.77$.
These properties are remarkably similar to the well-known binary evolution pulsator (BEP) reported by \citet{BEP}. That system has an orbital period of $15.9$~days and a pulsation period of $0.62$~days. Its expected bolometric luminosity is $\log(L/L_\odot)=1.55$–$1.75$. The two objects are therefore closely analogous. In both cases, the absolute magnitude predicted from the metallicity–magnitude relation is fainter than the observed value. For this reason, CV~Scl could be another BEP candidate.

However, in the Gaia non-single star catalogue the SB1 solution is flagged with an intermediate period-confidence (flag bit 8192), indicating that the orbital period may be spurious and that the solution requires independent confirmation.

To investigate this further, we searched the MAST archive\footnote{\url{https://mast.stsci.edu/portal/Mashup/Clients/Mast/Portal.html}} and found that the star has been observed by TESS \citep{Ricker15}. The data span $\sim700$ days. The observations consist of four windows of 20–30 days each, with high cadence ($\lesssim 1$ minute). We used these data to build an O–C diagram. We do not find evidence for phase modulation consistent with a binary orbit with a period close to 14 days.

At this stage, these results are preliminary and non conclusive. Future dedicated observations will be required to confirm or rule out this source as a BEP candidate. For the purposes of this work, this point is not critical. Even if CV~Scl is confirmed as a binary BEP source, it would not belong to the class of binary-formed RRLs investigated here (Section~\ref{sec:binmodel}).

\subsection{Full orbital solution: \Oone (V552~Car)} \label{app:V552Car}

O1 (also known as V552~Car) is the only RRL candidate in Gaia with a complete astrometric solution, complemented by radial-velocity information that allows a correction for perspective acceleration in the astrometric fit (see \citealp{GaiaDR3binary}). The inferred orbital period ($\approx 2200$~days) is broadly consistent with expectations for a binary-made RRL star considered in this work (Section~\ref{sec:binmodel}).
However, the classification of this object remains uncertain. While Gaia lists it as an RRc, the \textit{General Catalogue of Variable Stars} \citep{Samus17} classifies it as a BY~Draconis variable, i.e. a low-mass (K–M type) main-sequence star showing rotational photometric modulation induced by starspots. Its amplitude ($\approx 0.2$~mag in the $G$-band) and variability period ($\approx 0.3$~days) place it close to the boundary between these two variability classes.
Nevertheless, its apparent magnitude ($G\approx 12$~mag), reddening ($E_{B-V}=0.55$, corresponding to an extinction of $\approx 1.4$~mag), and parallax ($\approx 6.9$~mas; $d\approx 145$~pc) imply an absolute magnitude of $\approx 5$, too faint for an RRL \citep[see][for similar cases]{Ranaivomanana2025}. Spectroscopy further supports a non-RRL nature, indicating a temperature too low for an RRL ($\lesssim 5500$K), significant chromospheric activity, and the presence of lithium \citep{Marsden2009,Yamashita2022}. In addition, its position, distance, magnitude, and colours are consistent with membership in the young ($\approx 30$~Myr) open cluster IC2602 \citep{Jackson2020,Gutierrez2020,Nisak2022}. 
Overall, the available evidence strongly suggests that V552~Car is a young main-sequence or pre-main-sequence star whose variability is driven by rotational modulation, and that it was misclassified as an RRc in Gaia DR3.
The astrometric detection of binarity remains intriguing, as the star had not been recognised as a binary system prior to Gaia DR3. If confirmed, a young highly eccentric binary ($e\approx 0.8$) would represent an interesting target for follow-up observations, although a detailed analysis lies beyond the scope of this work.

\subsection{Non-linear proper motion, \Aone} \label{sec:aone}

Figure~\ref{fig:rrlbindr3lc} clearly shows that the pulsation period predicted for A1 in Gaia DR3 (0.36~days) is incorrect. Indeed, a Lomb–Scargle periodogram indicates that a period of $0.148$~days provides a significantly better description of the source variability (see bottom-right panel), in agreement with  \cite{Pranav26}. The corrected light curve suggests that the source is not an RRab, as reported in Gaia, but is instead more consistent with the nearly sinusoidal morphology expected for an RRc-type pulsator. However, the pronounced asymmetry between the minimum and maximum brightness (i.e. the deep dimming compared to the peak) suggests an eclipsing nature. This interpretation is further supported by the reduced scatter obtained when folding the light curve at twice the best-fitting period ($0.296$~days), which would correspond to the orbital period in the eclipsing-binary scenario.

\section{Binary properties sampled from the DPGMM model} \label{app:DPGMM}

\begin{figure*}
    \centering
    \includegraphics[width=0.8\linewidth]{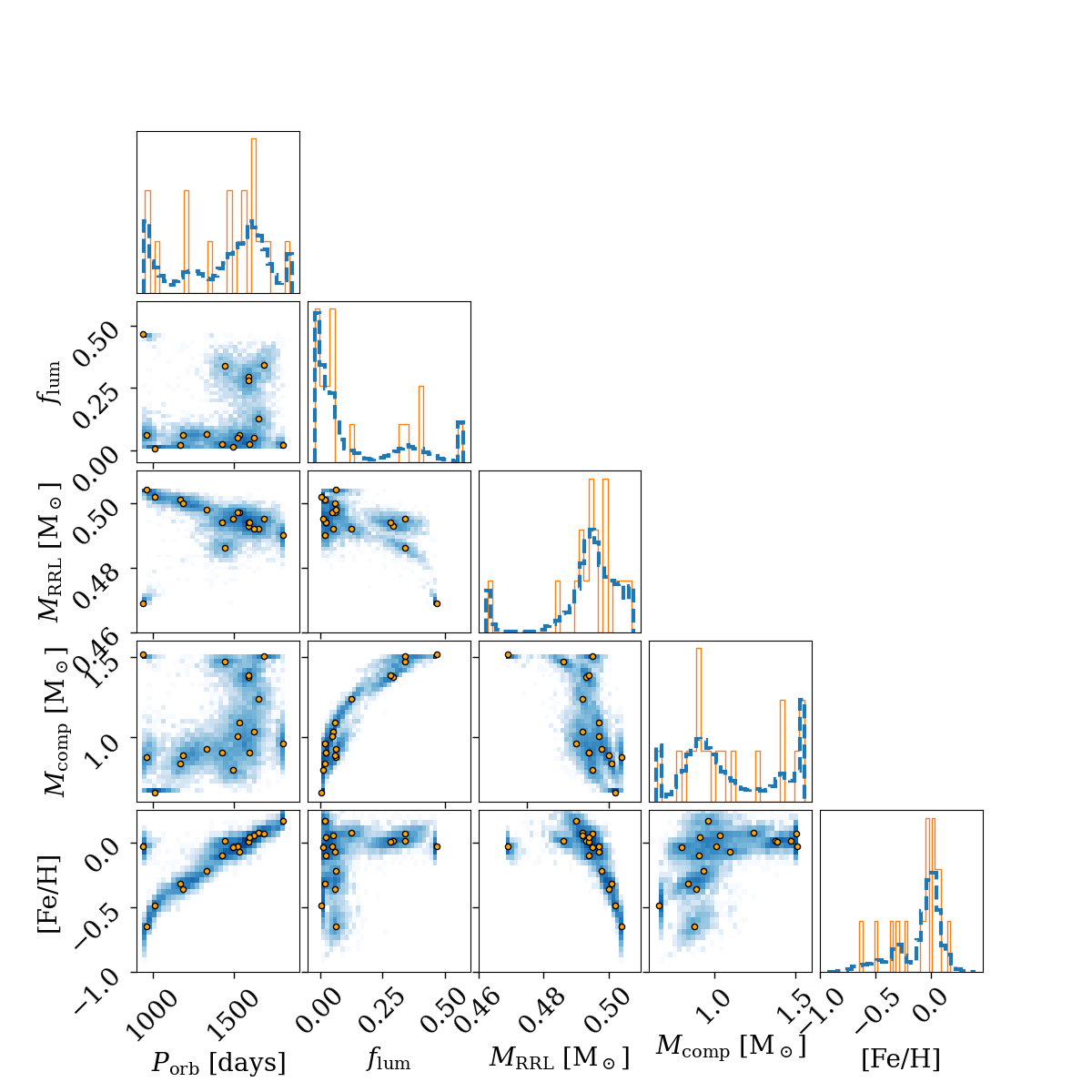}
    \caption{Corner plot showing  $10^4$ realisations drawn from the DPGMM model (blue colormaps and histograms) fitted on the RRL in binary sample from \cite{BI24} (orange points and histograms). The 1D histograms represents pdfs, while the blue colormap shows the number counts in a logarithmic scale. The sample variables are: the orbital period ($P_\mathrm{orb}$), the companion-RRL luminosity ratio in the Gaia $G$ band ($f_\mathrm{lum}$), the mass of the RRL ($M_\mathrm{RRL}$), the mass of the main sequence companion ($M_\mathrm{comp}$), and the metallicity ([Fe/H]). }
    \label{fig:DPGM}
\end{figure*}

Figure~\ref{fig:DPGM} shows the sampled orbital properties of binary-formed RRLs used in this work (Section~\ref{sec:binmodel}). These are drawn from a DPGMM fitted to the \citet{BI24} simulations, which assumes that the 22 binary-made RRLs represent a realisation of a parent probability density function (pdf). The resulting pdf fully accounts for the uncertainties arising from the finite sample size.

\section{\texttt{gaiamock} update:  heteroskedastic errors and variability-induced mover  effect and fit} \label{app:vim}

To account for epoch-to-epoch variations in astrometric uncertainties and for the variability-induced mover (VIM) effect, we modify the generation of astrometric data for both single and binary sources. In this framework, astrometric errors (which depend on the $G$-band magnitude) and the flux ratio are no longer fixed at their mean values, but vary at each epoch according to
$
f_\mathrm{var}(t) = f\,10^{0.4 dG(t)},
$
where $f$ is the flux ratio inferred from the mean magnitude and $dG(t)$ is the magnitude variation predicted by the variability model. This prescription naturally accounts for the VIM effect in the astrometric signal (see Equation~1 in \citealt{ElBadry24}).

In Gaia DR3, an astrometric model explicitly based on the VIM effect, referred to as the fixed-VIM (VIMF) model, is tested \citep{GaiaDR3binary}. This model assumes that the photocentric motion is dominated by variability-induced shifts, with orbital motion neglected or absorbed into the fitted parameters. We implement the VIMF model in \texttt{gaiamock} following the Gaia DR3 prescription. The astrometric displacement is modelled as
\begin{equation}
\eta(t) = \eta_\mathrm{5par}
+ D_\alpha \left( \frac{F_\mathrm{G,ref}}{F_\mathrm{G}(t)} - 1 \right) \sin \psi
+ D_\delta \left( \frac{F_\mathrm{G,ref}}{F_\mathrm{G}(t)} - 1 \right) \cos \psi ,
\label{eq:vimf}
\end{equation}
where $\eta_\mathrm{5par}$ is the standard 5-parameter astrometric model (Equation~3 in \citealt{ElBadry24}), $\psi$ is the Gaia scanning angle, and the vector parameter $D=(D_\alpha,D_\delta)$ introduces two additional free parameters describing the amplitude and direction of the VIM-induced displacement.

Adopting the mean $G$-band flux as reference, the term $F_\mathrm{G,ref}/F_\mathrm{G}(t)$ reduces to $10^{0.4 dG(t)}$. The per-transit $G$ magnitudes entering Equation~\ref{eq:vimf} are obtained by convolving the model light curve with the Gaia DR3 photometric uncertainties, computed using the relations from \citet{Riello21}. In the VIMF model, the effective uncertainties depend on both the astrometric and photometric errors, as well as on the fitted parameter $D$ (see Equation~18 in \citealt{GaiaDR3binary}). Following the Gaia pipeline, we therefore perform the fit iteratively until the best-fit parameters change by less than 1\% between successive iterations.

\section{Hierarchical Bayesian sparse Gaussian process FITC model} \label{app:fitc}

To estimate the dependence of the binary fraction on RRL metallicity within a hierarchical Bayesian framework, we adopt a flexible non-parametric model. In particular, we combine a logistic link function with a latent function described by a Gaussian Process (GP), as defined in Equation \ref{eq:fbin}.
Standard GP regression requires evaluating the covariance matrix for the full set of $N$ training points. In this case, the likelihood evaluation involves the inversion of an $N\times N$ covariance matrix, resulting in a computational cost scaling as $\mathcal{O}(N^3)$ and a memory requirement scaling as $\mathcal{O}(N^2)$.

To reduce this complexity, we use the Fully Independent Training Conditional (FITC) approximation \citep{Candela2005}, which introduces a set of $M$ inducing points. If $M<N$, FITC replaces the full GP covariance with a low-rank approximation in which correlations between training points are mediated by the inducing ones. In practice, the dense covariance is only computed between the training points (our RRL sample) and the inducing points, while the remaining small-scale residual variance is retained through an additional diagonal correction term. This yields a sparse approximation that preserves much of the flexibility of a full GP while reducing the computational scaling to $\mathcal{O}(NM^2)$, enabling efficient inference on moderately large samples.
In our specific case, $N=100$ and we adopt $M=15$, reducing the time complexity of likelihood evaluations by approximately two orders of magnitude (from $\mathcal{O}(10^6)$ to $\mathcal{O}(10^4)$ operations). With this configuration, full posterior sampling typically requires $\sim 2$–$4$ hours to reach convergence. A full GP analysis would instead require of order $\sim 10$ days, significantly limiting our ability to explore alternative model configurations.

Beyond computational convenience, the use of a sparse inducing-point GP is also consistent with the aim of this work, which is to constrain broad trends in the maximum allowed binary fraction as a function of metallicity, rather than producing highly local predictions at each individual metallicity value sampled by the analysed RRLs.

Table~\ref{tab:fitc} summarises the hierarchical Bayesian FITC-GP models implemented and sampled using \texttt{PyMC3} \citep{PYMC}. During sampling, the true metallicity of each RRL ($\mathrm{[Fe/H]}^\ast$) is treated as a latent variable and sampled from a Gaussian distribution centred on the photometric metallicity, with width given by its observational uncertainty.
To improve numerical stability, we standardise the metallicities using the mean $\langle \mathrm{[Fe/H]} \rangle$ and standard deviation $\sigma_{\mathrm{[Fe/H]}}$ of the sample (Equation \ref{eq:xfeh}).

During inference, the inducing-point covariance matrix $K_{MM}$ must be manipulated repeatedly  (e.g. to draw correlated latent variables). For numerical stability and computational efficiency, we factorise this matrix using a Cholesky decomposition,
\begin{equation}
K_{MM} = L_{MM}L_{MM}^{T},
\end{equation}
where $L_{MM}$ is a lower-triangular matrix. This decomposition avoids explicit matrix inversion, which is both slower and more prone to numerical instabilities. Moreover, it provides a convenient way to sample multivariate Gaussians.
We then adopt a non-centred parameterisation for the inducing-point latent variables. Rather than sampling correlated variables directly from
\begin{equation}
u \sim \mathcal{N}(0, K_{MM}),
\end{equation}
we sample independent standard normal variables,
\begin{equation}
u_\mathrm{aux} \sim \mathcal{N}(0, I),
\end{equation}
and construct
\begin{equation}
u = L_{MM}u_\mathrm{aux}.
\end{equation}
This transformation preserves the correct covariance ($\mathrm{Cov}(\mathbf{u}) = K_{MM}$) while reducing posterior correlations between the latent GP values and the GP hyperparameters. It therefore alleviates the ``funnel'' geometry that often appears in hierarchical Bayesian models due to the strong coupling between latent variables and their scale hyperparameters ($a_\mathrm{GP}$ and $\ell_\mathrm{GP}$ in our case). In particular, when a scale hyperparameter approaches small values, the latent distribution collapses into a narrow region of parameter space, producing strong curvature and correlations that can lead to slow mixing and convergence issues in HMC/NUTS sampling.
Although mathematically equivalent to the centred formulation, the non-centred parameterisation typically reduces posterior correlations, thereby improving numerical stability, sampling efficiency, and convergence. The same strategy is then applied to sample the latent values $f^\ast$ (see Table~\ref{tab:fitc}).

In this framework, the a posteriori probability, $D_i$, of detecting the $i$-th source as a binary RRLs in Gaia follows a Bernoulli distribution. 
Accordingly, the posterior distribution of the total number of binary detections, $P(N_\mathrm{D})$ is obtained by summing all independent Bernoulli trials, resulting in a Poisson Binomial distribution.

Considering the detectability posterior for each star (Equation \ref{eq:detectability}) and conditioning P($N_\mathrm{D}$) on a specific binary fraction, $f_\mathrm{bin}$, the mean of the conditioned posterior for a sample of $N$ stars is \citep{Gelman2014}: 
\begin{equation}
\langle N_\mathrm{D}  \rangle 
= f_\mathrm{bin}  \sum_{i=1}^{N}  \langle \mathcal{O} \rangle _i
= N f_\mathrm{bin}  \langle \mathcal{O} \rangle_\mathrm{s},
\label{eq:Ndet}
\end{equation}
where $\langle O \rangle_i$ is the  is the mean detectability for the $i$-th star in the sample, and $\langle \mathcal{O} \rangle_\mathrm{s}$ denotes the mean of the detectability considering the whole sample.
The variance is instead  given by
\begin{equation}
\sigma^2_{N_\mathrm{D}}
= \langle N_\mathrm{D} \rangle -
		 f_\mathrm{bin}^2 \sum^N_i \langle \mathcal{O} \rangle^2 _i. 
= \langle N_\mathrm{D} \rangle -
		  N f_\mathrm{bin}^2  \langle  \langle \mathcal{O} \rangle^2_i  \rangle_\mathrm{s}, 
\label{eq:Nvariance}    
\end{equation}
where $\langle \langle \mathcal{O} \rangle_i^2 \rangle_\mathrm{s}$ denotes the sample average of the squared mean detectabilities.
Equation \ref{eq:Nvariance} combines the variance of the Poisson-Binomial with the variance of the detectability posterior by using the law of total variance\footnote{Given two stochastic variables on the same probability space  (in our case $N_\mathrm{D}$ and $\mathcal{O}$), the total variance on  one variable ($N_\mathrm{D}$) is $\sigma^2_{N_\mathrm{D}}= \mathrm{Var}({N_\mathrm{D}}) =\langle \mathrm{Var}({N_\mathrm{D}|\mathcal{O})} \rangle   +  \mathrm{Var}(\langle N_\mathrm{D}|\mathcal{O} \rangle) = \langle \sum_i (1-f_\mathrm{bin} \mathcal{O}_i) f_\mathrm{bin} \mathcal{O}_i \rangle  + f^2_\mathrm{bin} \sum_i \mathrm{Var}(\mathcal{O}_i)$, and $\mathrm{Var}(\mathcal{O}_i)=\langle \mathcal{O}^2  \rangle_i - \langle \mathcal{O}  \rangle^2_i$}.

\begin{table*}
\centering
\caption{Hierarchical sparse-GP FITC model for $f_\mathrm{bin}(Z)$ inference and Gaia DR4 predictions}
\label{tab:model}
\small
\begin{tabular}{ll}
\hline\hline
\textbf{Component} & \textbf{Specification} \\
\hline

\textit{GP hyperparameters} &
$\log a_\mathrm{GP}^2 \sim \mathrm{Uniform}(0,2), \quad
\log \ell_\mathrm{GP} \sim \mathrm{Uniform}(-3,1)$ \\[2mm]

\textit{Link function  hyperparameters} &
$k \sim \mathrm{HalfNormal}(2)$, \;
$f_0 \sim \mathcal{N}(0,2)$ \\[2mm]

\textit{Latent metallicities} &
$\displaystyle \mathrm{[Fe/H]}^\ast \sim  \mathcal{N}(\mathrm{[Fe/H]},\,\delta_{\mathrm{[Fe/H]}})$ \\[2mm]

\textit{Rescaled metallicities} &
$\displaystyle x = \left( \mathrm{[Fe/H]}^\ast - \langle \mathrm{[Fe/H]} \rangle \right) \sigma^{-1}_{\mathrm{[Fe/H]}}  $ \\[2mm]

\textit{Kernel} &
Squared-exponential: $k_\theta(x,x') = a_\mathrm{GP}^2 \exp[-(x-x')^2/(2\ell^2_\mathrm{GP})]$ \\[2mm]

\textit{GP Inducing-point, m,  correlation matrix} &
$K_{MM} = k_\theta(M,M) + 10^{-6}I$, \;
$u_{\rm aux} \sim \mathcal{N}(0,I)$, \;
$u = L_{MM} u_{\rm aux}$ \\[2mm]

\textit{GP Mean at data} &
$\mu_f = K_{X M} K_{MM}^{-1} u$ \\[2mm]

\textit{FITC diagonal correction} &
$Q_{XX} = K_{X M} K_{MM}^{-1}K_{M X}$, \;
$\Lambda = \mathrm{diag}(K_{XX}-Q_{XX})$ \\[2mm]

\textit{Latent GP values} &
$f_{\rm aux} \sim \mathcal{N}(0,I)$, \;
$f^\ast(x) = \mu_{f} + \sqrt{\Lambda}\, f_{{\rm aux}}$ \\[2mm]

\textit{Binary-fraction link} &
$f_{\rm bin}(x) = \left( 1+e^{-k(f^\ast(x)-f_0)} \right)^{-1}$ \\[2mm]

\textit{Detectability (DR3)} &
$O_i \sim \mathrm{Beta}(1 + N_{{\rm b},i},\, 1 + N_{{\rm r},i}-N_{{\rm b},i})$ \\[2mm]

\textit{DR3 predictions} &
$D_i \sim \mathrm{Bernoulli}(f_{\rm bin}(x_i) O_i)$, \;
$N_{\rm D} = \sum_i D_i$ \\[2mm]

\textit{DR4  detectability} &
$O_i^{\rm DR4} \sim \mathrm{Beta}(1+N^{\rm DR4}_{{\rm b},i},\,
1+N^{\rm DR4}_{{\rm r},i}-N^{\rm DR4}_{{\rm b},i})$ \\[2mm]

\textit{DR4 predictions} &
$D_i^{\rm DR4} \sim \mathrm{Bernoulli}(f_{\rm bin}(x_i) O_i^{\rm DR4})$, \;
$N_{\rm D}^{\rm DR4} = \sum_i D_i^{\rm DR4}$ \\
\hline
\end{tabular}
\begin{tablenotes}
\small
\item Notes. The training points (rescaled metallicities of the analysed RRLs) are denoted by $x \in X$, while the inducing points are $m \in M$. The matrix $L_{MM}$ represents the upper-triangular matrix from the Cholesky decomposition of $K_{MM}$. The auxiliary variables $u_\mathrm{aux}$ and $f_\mathrm{aux}$ are used to implement the non-centred parametrisation of the Gaussian distribution. The index $i$ refers to an individual source in the sample, while $N_{{\rm b},i}$ represents the number of times \texttt{gaiamock} predicts a positive binary detection out of $N_{{\rm r},i}$ realisations for a given source (see Section~\ref{sec:detectability_definition}). Finally, $N_{{\rm D}}$ denotes the distribution of detections within the sample.
\end{tablenotes}
\label{tab:fitc}
\end{table*}

\section{Impact of mass and angular-momentum loss on final orbital periods} \label{app:angmom}

\begin{figure}
    \centering
    \includegraphics[width=\linewidth]{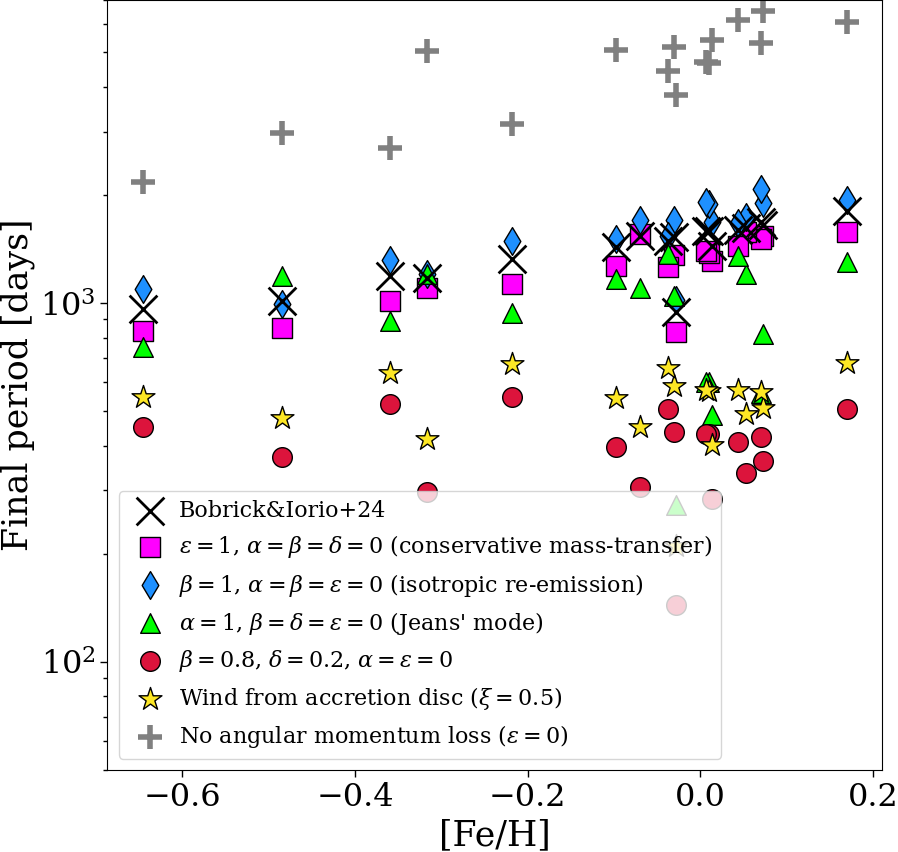}
    \caption{Final post-mass-transfer orbital periods estimated from Equation~\ref{eq:angmom}, using the initial periods, initial masses, and final RRL masses from the binary-formation models of \cite{BI24} (see Section~\ref{sec:binmodel} and Appendix~\ref{app:DPGMM}). Different symbols indicate different mass-loss prescriptions: fully conservative transfer (magenta squares), isotropic re-emission (blue diamonds), Jeans fast-wind mode (green triangles), a mixed case with 80\% isotropic re-emission plus 20\% mass loss through a circumbinary disc (red circles), and a mixed case with 50\% isotropic re-emission and 50\% mass loss from an accretion disc around the accretor (yellow stars, \citealt{Gallegos24}). The grey plus symbols indicate the upper limit on the final orbital period obtained from a model assuming non-conservative mass transfer with no loss of orbital angular momentum.
    Black diagonal crosses show the final orbital periods obtained directly from the detailed binary simulations of \cite{BI24}.}
    \label{fig:angmom}
\end{figure}

During a mass-transfer episode in a binary system, the fraction of mass that is not accreted by the companion escapes from the system, carrying away orbital angular momentum. This loss of angular momentum, together with the change in the stellar masses, alters the orbital period (see \citealt{Tauris23} and references therein for a comprehensive overview).
In the standard formalism, the mass lost by the donor and not accreted onto the companion can leave the system through different channels:
as a fast isotropic wind from the donor (Jeans mode),
as isotropic re-emission from the vicinity of the accretor, or
through a circumbinary disc formed by material escaping via the outer Lagrangian points (L2/L3).

In reality, these channels may operate simultaneously during a mass-transfer episode. Following \citet{Tauris23}, we parameterise the fractions of mass lost through these channels as $\alpha$ (donor wind), $\beta$ (re-emission from the accretor), and $\delta$ (circumbinary disc). The fraction accreted by the companion is then
$
\epsilon = 1 - \alpha - \beta - \delta .
$
Each mass-loss mode removes a different amount of specific angular momentum from the system, leading to different final orbital separations and periods. \citet{Tauris06} derived a general analytic expression for the change in orbital separation of the form:
\begin{equation}
\frac{a}{a_0}
= \left(\frac{q}{q_0}\right)^{2\left( \alpha + 1.5 \delta -1 \right)} \left(\frac{q+1}{q_0+1}\right)^{\frac{-\alpha - \beta + \delta}{1-\epsilon}} 
\left( \frac{\epsilon q +1 }{\epsilon q_0 +1} \right)^{3+2\frac{\alpha \epsilon^2 + \beta + 1.5\delta(1-\epsilon)^2}{\epsilon(1-\epsilon)}},
\label{eq:angmom}
\end{equation}
where $q$ and $q_0$ denote the donor-to-accretor mass ratio at the end and at the onset of the mass-transfer episode, respectively. The specific form of Equation~\ref{eq:angmom} assumes, following stability arguments, that the circumbinary disc has a characteristic semi-major axis equal to 2.25 times the binary orbital separation \citep{Soberman97}.

The binary-formation model of \citet{BI24} analysed in this work is based on simulations in which the mass transfer results to be highly non-conservative, assuming that  the  mass is lost through isotropic re-emission from the accretor ($\beta = 1$, $\epsilon = 0$).
To explore how alternative mass-loss assumptions affect the final orbital separation and period, we use Equation~\ref{eq:angmom}. Starting from the initial mass ratio $q_0$, the initial orbital period of the binary-formed RRLs, and the final RRL masses (orange points in Figure~\ref{fig:DPGM}), we compute the corresponding final orbital period for different choices of $\alpha$, $\beta$, and $\delta$. The final companion mass, and thus the final mass ratio $q$, are determined self-consistently from the assumed accreted fraction $\epsilon$.
Equation~\ref{eq:angmom} diverges for $\epsilon = 1$ (fully conservative mass transfer) and $\epsilon = 0$. In the conservative case, we instead adopt the standard expression for conservative mass transfer (Equation~4.40 in \citealt{Tauris23}):
\begin{equation}
\frac{a}{a_0} = \left( \frac{q}{q_0} \right)^{-2} \left( \frac{q+1}{q_0+1} \right)^4.
\end{equation}
In the fully non-conservative limit ($\epsilon = 0$), we approximate $\epsilon$ with a small value ($10^{-10}$) and rescale $\alpha$, $\beta$, and $\delta$ by a factor $(1 - 10^{-10})$.
In addition, we explore an extension of the isotropic re-emission model in which part of the transferred mass is expelled as a wind from an accretion disc around the accretor (see e.g. \citealt{Bobrick17}). In this case, the ratio between final and initial orbital separation follows Equation~13 of \cite{Gallegos24}, which assumes no net accretion onto the companion and depends on the fraction of mass lost from the accretion disc, $\xi$, and on the outer radius of the disc (adopted by default as 75\% of the accretor Roche-lobe radius). For $\xi=0$, this prescription reduces to the standard isotropic re-emission model ($\beta=1$, $\epsilon=0$).
Finally, to set an upper limit on the possible orbital widening, we consider an extreme scenario in which no orbital angular momentum is lost from the system, even if the mass transfer is non-conservative. This case can be derived from Equation \ref{eq:angmom} assuming $\alpha=\beta=\delta=0$.
Conceptually, this limiting case corresponds to mass being lost isotropically from the system barycentre, so that it carries away no net orbital angular momentum. Although idealised, it may approximate situations in which the accretor is significantly more massive than the donor and lies very close to the system barycentre, as in a black hole–low-mass star binary.

Figure~\ref{fig:angmom} compares the predicted final orbital periods obtained under different mass- and angular-momentum loss prescriptions with those from the detailed binary simulations. As expected, the isotropic re-emission model closely reproduces the simulations, since it matches their underlying assumptions, although it neglects additional effects such as prior wind mass loss and tidal interactions. Excluding the extreme case of zero angular-momentum loss, isotropic re-emission produces the longest post-mass-transfer periods for the binary configurations considered here.
Fully conservative transfer and Jeans’ wind mode give broadly similar results, although the latter predicts some systems at near-solar metallicity with periods in the range $\sim200$–$500$ days. Even a modest fraction of mass lost through a circumbinary disc ($\sim20\%$) significantly shrinks the orbit, shifting the entire population to $\sim200$–$500$ days and erasing the metallicity–period correlation. Larger values of $\delta$ lead to progressively tighter systems, reaching $\sim1$–$10$ days for $\delta=1$.
Mass loss through an accretion disc spans from the isotropic re-emission limit ($\xi=0$) to a distribution nearly identical to the 20\% circumbinary case ($\xi=1$). For an intermediate value, $\xi=0.5$, most final periods fall in the range $\sim400$–$600$ days, again with no clear metallicity–period trend.
Finally, the extreme case of non-conservative transfer with no angular-momentum loss sets an upper bound, showing that, in principle, final periods as long as $\sim2000$–$7000$ days could be accommodated.

These differences can be understood in terms of how efficiently each channel removes angular momentum. In binary-formed RRL systems, the accretor is more massive than the donor during most of the mass-transfer phase and therefore carries the lowest specific angular momentum in the system. Mass lost from its vicinity (as in isotropic re-emission) thus extracts relatively little angular momentum, leading to wider final orbits. By contrast, mass lost from the donor, or especially from a circumbinary disc, or an accretion disc removes angular momentum much more efficiently, producing shorter periods. In the fully conservative case no angular momentum is lost, but internal mass redistribution alone still leads to slightly shorter final periods.

This analysis is based solely on the results from Equation~\ref{eq:angmom} and is intended only as a qualitative indication of how different mass-loss assumptions may influence the final orbital separation.
It does not account for how mass loss and mass transfer may affect the subsequent binary evolution beyond their impact on the orbital separation. For example, an enhancement of the mass-loss rate could lead to complete envelope stripping before the helium flash or trigger unstable mass transfer. Conversely, mass loss may cease earlier if the evolution of the Roche-lobe radius causes the donor to detach.

\section{Alternative RR Lyrae samples} \label{app:alternative}

\begin{figure*}
    \centering
    \includegraphics[width=0.8\linewidth]{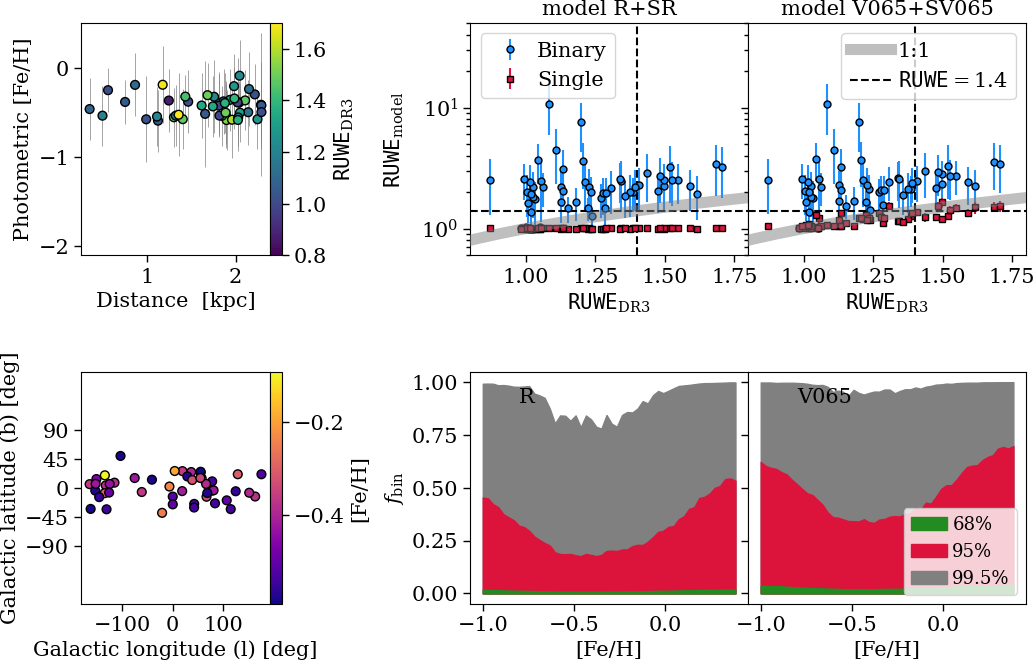}
    \caption{Summary plot for the alternative RRL sample  \texttt{Plx04} (see text). The two left panels show the distribution of photometric metallicity as a function of distance and the spatial distribution in Galactic coordinates. The two panels in the top-right reproduce the analysis shown in Figure~\ref{fig:ruwehist}, while the two bottom-right panels correspond to the results presented in Figure~\ref{fig:fbin}. }
    \label{fig:plx04}
\end{figure*}

\begin{figure*}
    \centering
    \includegraphics[width=0.8\linewidth]{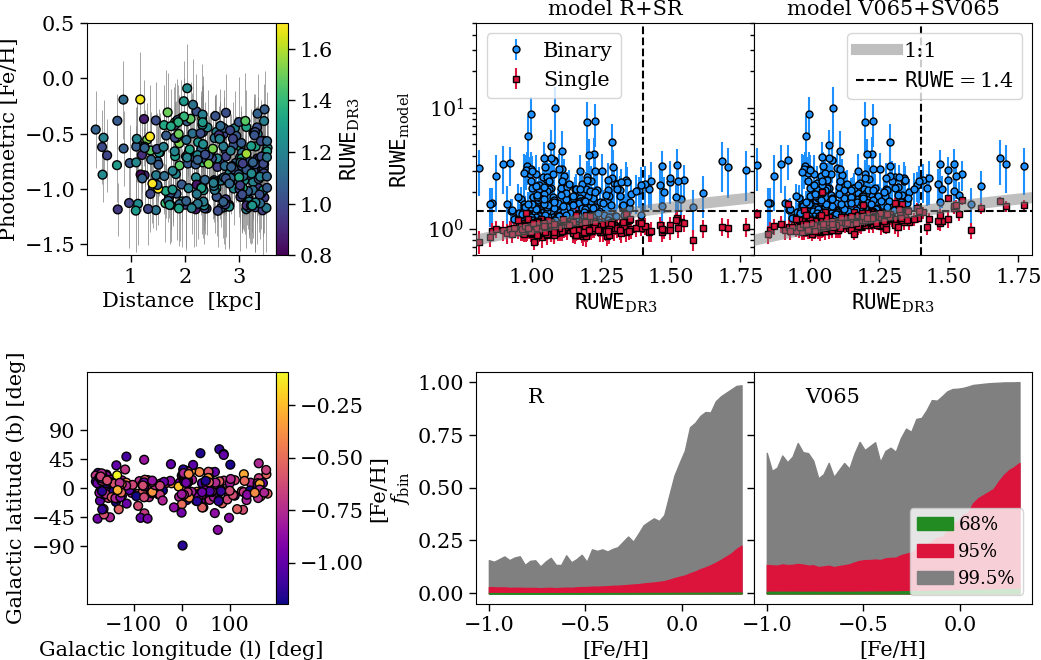}
    \caption{Same as Figure \ref{fig:plx04},  but for the alternative sample \texttt{ThinDisc} (see text).}
    \label{fig:thindisc}
\end{figure*}

\begin{figure*}
    \centering
    \includegraphics[width=0.8\linewidth]{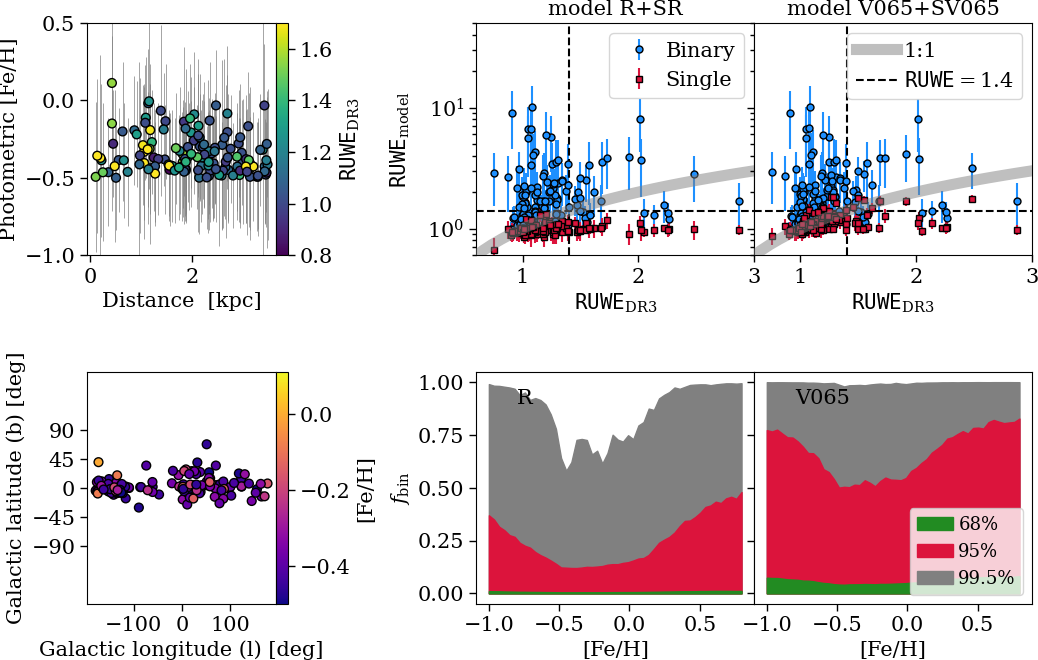}
    \caption{Same as Figure \ref{fig:plx04},  but for the alternative sample \texttt{RRabHighmet} (see text).}
    \label{fig:feh05}
\end{figure*}

\begin{figure*}
    \centering
    \includegraphics[width=0.8\linewidth]{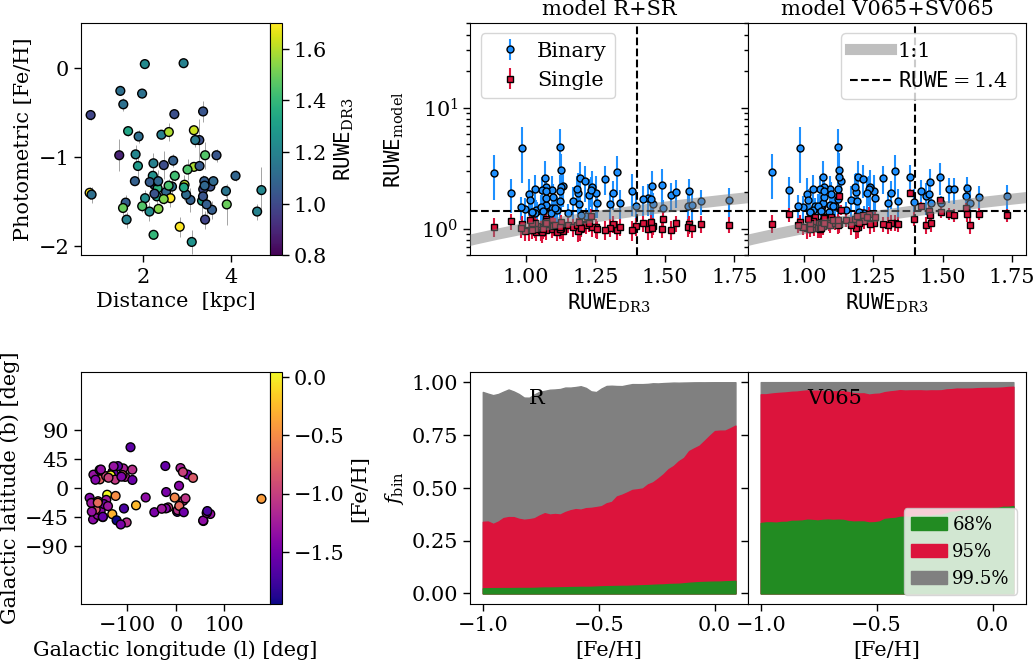}
    \caption{Same as Figure \ref{fig:plx04}, but for the alternative sample \texttt{DOrazi24} (see text).}
    \label{fig:dorazi24}
\end{figure*}

We test our analysis using four additional RRL samples, labelled \texttt{Plx04},\texttt{ThinDisc}, \texttt{RRabHighmet}, and \texttt{DOrazi24}, as alternatives to the one presented in the main text (Section~\ref{sec:sample}).

The \texttt{Plx04} sample is a subsample of the main one, obtained by applying a stricter parallax cut ($\varpi > 0.4$ mas). This subsample contains 48 stars.
Figure~\ref{fig:plx04} shows the sky distribution, distances, and metallicities of the sample, together with a summary of the data--model \ruwe\ comparison and the $f_\mathrm{bin}$ posterior for the R and V065 models (see Table~\ref{tab:models}).
On average, this subsample has a larger mean detectability (e.g.\ $\approx$28\% for the reference model, $\approx$20\% for the V065 model). Indeed, in the original sample, most sources ($\approx$70\%) with very low detectability ($\lessapprox 0.1$) belong to the population of stars with $\varpi < 0.4$ mas.
However, given the smaller number of stars, we cannot rule out a high binary fraction ($f_\mathrm{bin} \approx 0.7$) in any of the tested models.

In the \texttt{ThinDisc} sample, we relax the metallicity cut to a less stringent condition, $\mathrm{[Fe/H]} > -1.2$. At the same time, we enforce a stricter parallax signal-to-noise cut ($\varpi/\delta_{\varpi} > 5$) and introduce a kinematic selection, retaining only sources with a posterior probability of belonging to the thin disc greater than 0.7, based on the analysis of \cite{IB21}.
The final sample contains 286 objects; its distribution and results are summarised in Figure~\ref{fig:thindisc}.
Given the larger number of objects compared to the main sample, the constraints on $f_\mathrm{bin}$ are correspondingly tighter. 
When chromaticity bias is included, $f_\mathrm{bin} \approx 0.5$ cannot be excluded for [Fe/H]$\gtrsim-0.3$.
Although this value is lower than that obtained with the main sample, we do not interpret it as being in strong tension with the predictions of the binary-formation models, given the expected larger contamination by metal-poor RRLs from the thick disc and halo (see, e.g., \citealt{Abdollahi2025}).
Concerning \ruwe, $\approx 70\%$ of the sample is consistent within $3\sigma$ with the predictions of the binary models, with a systematic offset peaking at $\approx 0.5$. Similar to the main sample, the single-star model accounting for chromaticity bias reduces the average offset to approximately zero, with a mild dispersion of 0.1–0.2, consistent with the estimated uncertainties.

For the \texttt{RRabHighmet} sample, we do not adopt the subsample from \cite{Zhang25} and therefore do not include the astrometric quality cuts from \citet{IB21}. Instead, we select sources based solely on a parallax signal-to-noise ratio greater than 5, a metallicity cut of $\mathrm{[Fe/H]} > -0.5$, and we restrict the sample to RRab stars. The final sample contains 138 objects.
Figure~\ref{fig:feh05} shows that the results obtained with this sample are consistent with those of the main sample (Section \ref{sec:sample}). In particular, the posterior on the binary fraction is consistent with high values ($f_\mathrm{bin} \approx 0.75$) even for the reference model.
The \ruwe\ offset is slightly lower than in the \texttt{ThinDisc} sample, with about 80\% of the sample consistent with the binary-model predictions. However, this sample is likely more affected by poorly modelled astrometric outliers or contaminants, since the sources with the highest observed \ruwe\ ($>2$) are predicted by the models, both in the binary scenario and in the single-star case with chromaticity bias, to have substantially lower \ruwe\ values.

The \texttt{DOrazi24} sample includes 72 sources from the cross-match between Gaia DR3 and the high-resolution spectroscopic sample of \cite{DOrazi24}. The metallicity estimates in this sample are significantly more robust than the photometric metallicities used in our work; however, only eight objects have $\mathrm{[Fe/H]} > -0.6$. Therefore, the vast majority of the RRLs in the sample are likely associated with the old components of thick disc and stellar halo.  The results are shown in 
Figure~\ref{fig:dorazi24}.
Given these small numbers, the $f_\mathrm{bin}$ posterior is almost unconstrained, 
especially when chromaticity bias is included, and the results remain consistent with those obtained for the main sample.

\end{appendix}

\end{document}